\documentclass[a4paper,twoside]{article}
\newcommand{\affil}[1]{$^{\rm #1}$}
\usepackage[authoryear]{natbib}
\bibpunct{(}{)}{;}{a}{}{,}
\usepackage{graphicx}
\usepackage{url}
\usepackage{fixltx2e}
\date{} 
\newcommand{\kms}{\mbox{km\,s$^{-1}$}}
\newcommand{\cms}{\mbox{cm$^{2}$}}
\newcommand{\mab}{\mbox{$m_{AB}$}}
\newcommand{\ergs}{ergs\,cm$^{-2}$\,s$^{-1}$\,sr$^{-1}$}
\newcommand{\degsq}{\mbox{deg$^{2}$}}
\newcommand{\degs}{\mbox{$^{\circ}$}}
\newcommand{\msun}{\mbox{M$_{\odot}$}}
\newcommand{\mjup}{\mbox{M$_{J}$}}
\newcommand{\si}{\mbox{$\sim$}}
\newcommand{\mm}{\mbox{$\mu$m}}
\newcommand{\mj}{\mbox{$\mu$Jy}}

\newcommand{\vb}{\emph{V}-band}

\newcommand{\rb}{\emph{r}-band}

\newcommand{\kb}{\emph{K}-band}
\newcommand{\kd}{\mbox{\emph{K}$_{d}$}}
\newcommand{\ks}{\mbox{\emph{K}$_{s}$}}

\newcommand{\lb}{\emph{L}-band}
\newcommand{\ls}{\emph{L}$_{s}$}
\newcommand{\mb}{\emph{M}-band}
\newcommand{\nb}{\emph{N}-band}

\newcommand{\Hy}{H$_{2}$}
\newcommand{\ocen}{$\omega$ Cen}
\newcommand{\ocenn}{$\omega$ Centauri}
\newcommand{\bcep}{$\beta$ Cephei}

\title{\large\bf\flushleft The Science Case for PILOT III: the Nearby Universe}
\author{\parbox{\textwidth}{\flushleft
\vspace{-0.5cm}
{\it J.S.~Lawrence\affil{\,A,M}, M.C.B.~Ashley\affil{\,A}, J.~Bailey\affil{\,A}, D.~Barrado~y~Navascues\affil{\,B},
T.~Bedding\affil{\,C}, J.~Bland-Hawthorn\affil{\,C}, I.~Bond\affil{\,D}, H.~Bruntt\affil{\,C}, M.G.~Burton\affil{\,A}, M.-R.~Cioni\affil{\,E},
C.~Eiroa\affil{\,F}, N.~Epchtein\affil{\,G}, L.~Kiss\affil{\,C}, P.O.~Lagage\affil{\,H},
V.~Minier\affil{\,H}, A.~Mora\affil{\,F}, K.~Olsen\affil{\,I}, P.~Persi\affil{\,J},
W.~Saunders\affil{\,K}, D.~Stello\affil{\,C}, J.W.V.~Storey\affil{\,A}, C.~Tinney\affil{\,A}, and P.~Yock\affil{\,L}}\\
\vspace{0.4cm}
{\small \affil{A}\,School of Physics, University of New South Wales, NSW 2052, Australia}\\
{\small \affil{B}\,Laboratorio de Astrof\'{\i}sca Espacial y F\'{\i}sica Fundamental (INTA), Madrid 28080, Spain}\\
{\small \affil{C}\,Institute of Astronomy, School of Physics, University of Sydney, NSW 2006, Australia}\\
{\small \affil{D}\,Massey University, Auckland 0745, New Zealand}\\
{\small \affil{E}\,Centre for Astrophysics Research, University of Hertfordshire, Hatfield AL10 9AB, UK}\\
{\small \affil{F}\,Universidad Aut\'{o}noma de Madrid, C-XI, Madrid 28049, Spain}\\
{\small \affil{G}\,CNRS-Fizeau/UNSA, Nice 06108, France}\\
{\small \affil{H}\,Service d$'$Astrophysique, CEA Saclay, Saclay 91191, France}\\
{\small \affil{I}\,Kitt Peak National Observatory, National Optical Astronomy Observatory, Tucson, Arizona 85719, USA}\\
{\small \affil{J}\,Istituto Astrofisica Spaziale e Fisica Cosmica/INAF, Roma 00100, Italy}\\
{\small \affil{K}\,Anglo-Australian Observatory, NSW 1710, Australia}\\
{\small \affil{L}\,University of Auckland, Auckland 1142, New Zealand}\\
{\small \affil{M}\,JSL now at Department of Physics and Electronic Engineering, Macquarie University, NSW 2109, Australia; and Anglo-Australian Observatory, NSW 1710, Australia; Email: jsl@physics.mq.edu.au}}}
\begin{document}
\twocolumn[
\begin{changemargin}{.8cm}{.5cm}
\begin{minipage}{.9\textwidth}
\vspace{-1cm} \maketitle
%
%
\small{\bf Abstract:} PILOT (the Pathfinder for an International Large Optical Telescope)
is a proposed 2.5~m optical/infrared telescope to be located at Dome~C on the Antarctic
plateau. The atmospheric conditions at Dome~C deliver a high sensitivity, high
photometric precision, wide-field, high spatial resolution, and high-cadence imaging
capability to the PILOT telescope. These capabilities enable a unique scientific
potential for PILOT, which is addressed in this series of papers. The current paper
presents a series of projects dealing with the nearby Universe that have been identified
as key science drivers for the PILOT facility. Several projects are proposed that examine stellar populations in nearby galaxies and stellar clusters in order to gain insight into the formation and evolution processes of galaxies and stars. A series of projects will
investigate the molecular phase of the Galaxy and explore the ecology of star formation,
and investigate the formation processes of stellar and planetary systems. Three projects
in the field of exoplanet science are proposed: a search for free-floating low-mass
planets and dwarfs, a program of follow-up observations of gravitational microlensing
events, and a study of infrared light-curves for previously discovered exoplanets. Three
projects are also proposed in the field of planetary and space science: optical and
near-infrared studies aimed at characterising planetary atmospheres, a study of coronal
mass ejections from the Sun, and a monitoring program searching for small-scale Low Earth Orbit satellite debris items.


\medskip
\medskip
\end{minipage}
\end{changemargin}
] \small

\section{Introduction}

PILOT (Pathfinder for an International Large Optical Telescope) is proposed as a high
spatial resolution wide-field telescope with an optical design and an instrument suite
that are matched to the Dome~C atmospheric conditions. These conditions have been shown
to offer high infrared sensitivity, due to the low atmospheric thermal emission and low
water-vapour column-density \citep{Lawrence_04,Walden_e_05,Tomasi_e_06}; high spatial
resolution and high photometric precision, due to the unique atmospheric turbulence
structure above the site \citep{Lawrence_e_04,Agabi_e_06,Kenyon_e_06,Trinquet_e_08}; and
a high cadence, due to the high latitude of the site and the high percentage of
cloud-free conditions \citep{Kenyon_Storey_06,Mosser_Aristidi_07}.

The scientific justification for the PILOT telescope has evolved from earlier work
\citep{Burton_e_94,Burton_e_01,Burton_e_05} in parallel with the telescope and instrument
suite design \citep{Saunders_e_08a,Saunders_e_08b}. The PILOT science case is presented
here in a series of three papers. Paper I \citep{Lawrence_e_09a} gives a summary of the
science case, and an overview of the project (including the telescope design, expected
performance, and observing strategies). Paper II \citep{Lawrence_e_09b} presents a series
of science projects for the PILOT facility that are aimed at observing and understanding
the distant Universe (i.e., beyond a redshift of $z\approx1$). The current paper
discusses PILOT science projects dealing with the nearby Universe (i.e., the Solar
System, the Milky Way, and nearby galaxies).

The baseline optical design for PILOT, described in \citet{Saunders_e_08a}, comprises a
2.5~m Ritchey-Chretien telescope with an f/10 overall focal ratio. The telescope is
housed in a temperature- and humidity-controlled dome that is mounted on top of a \si30 m
high tower in order to elevate the main mirror above the majority of the intense
ground-layer turbulence. A fast tip-tilt secondary mirror is used for guiding and to
remove residual boundary-layer turbulence and tower wind-shake. As detailed in Paper I
\citep[see also][]{Saunders_e_08b} the baseline instrument suite for PILOT includes the
PILOT VISible Camera (PVISC), a wide-field ($40'\times40'$) optical imaging camera with a
spatial resolution of \si0.3$''$ over the wavelength range 0.4--1~\mm; the PILOT
Near-InfraRed Camera (PNIRC), a wide-field ($10'\times10'$) near-infrared camera
achieving a similar spatial resolution over the wavelength range 1--5~\mm; the PILOT
Mid-InfraRed Imaging Spectrometer (PMIRIS), a wide-field ($14'\times14'$) mid-infrared
instrument operating from 7--40~\mm, with several spectral-resolution modes; and the
PILOT Lucky Imaging Camera (PLIC), a fast optical camera for diffraction-limited imaging
over relatively small-fields ($0.5'\times0.5'$) in the visible.

The ``nearby Universe" science cases described in this paper are divided into four key
themes: \emph{stellar properties and populations}, \emph{star and planet formation},
\emph{exoplanet science}, and \emph{solar system and space science}. The projects proposed under these themes have been identified to take advantage of the unique discovery space of the PILOT telescope that is enabled by the Dome~C site conditions and the telescope design. For each project, the context and impact is discussed, along with the capabilities of competing (and synergistic) facilities. Additionally, the observational requirements are identified. It is shown that the observational requirements for these projects can either only be achieved with PILOT or can be achieved with other facilities but with greatly reduced efficiency. While many of the science projects described here, and in Paper II) require large amounts of observing time (several seasons in some cases), the observing strategies proposed in Paper I should allow the majority of these science projects to be accomplished within the proposed ten-year lifetime of the PILOT facility.

The \emph{stellar properties and populations} theme (Section 2) links several projects
that examine nearby galaxies and stellar clusters. The first project involves a study of
disc galaxies, in the optical and infrared, out to the edge of the Local Group. This
study aims to examine the relationship between stellar populations, star formation
history, and the local environment. A second project aims to examine the stellar content
of the outer regions of (nearer) satellite galaxies over wider fields, to gain insight
into the processes of galaxy formation and evolution. A project to image the Magellanic
Clouds to a greater depth with higher spatial resolution than existing surveys aims to
understand star formation processes and extreme populations of AGB stars. Finally, a
project involving long time-series optical observations of nearby globular and open
clusters aims to study age-metallicity relationships, test various predictions of stellar
astrophysics, and improve our understanding of the physics of massive stars.

The \emph{star and planet formation} theme (Section 3) links several projects that deal
with the early evolutionary stages of stellar and planetary systems. A wide-area survey
of the Milky Way Central Molecular Zone in the mid-infrared lines of molecular hydrogen
is proposed in order to understand the life history of the molecular phase of the Galaxy.
A survey of the Chamaeleon dark clouds complex at several lines in the mid-infrared aims
to explore planet formation via the identification of circumstellar discs around young
low-mass stellar objects and young brown dwarfs. We also discuss here the potential to
investigate the onset of planet formation in protoplanetary discs via the mid-infrared
spectral signatures of cyrstalline silicates, and the potential to investigate the very
early stages of star formation via the mid-infrared photometric characterisation of
embedded young stellar objects.

Three \emph{exoplanet science} studies are proposed in Section 4. The first project
involves a near-infrared search for planets ``free-floating" in nearby star clusters,
with the aim of directly detecting objects down to a few Jupiter masses. The second
project involves a program for optical follow-up of gravitational microlensing candidates
based on alerts from dedicated survey telescope networks; this is aimed in particular at
firmly determining the abundance of ice-giant planets. Finally, a project to obtain
high-precision near- and mid-infrared photometric light-curves of previously discovered
exoplanets aims to characterise the atmospheric properties of a large number of hot
Jupiters.

The \emph{solar system and space science} theme (Section 5) deals with several Solar System projects. The first project will investigate atmospheric and surface properties of Mars and Venus via high spatial resolution imaging in the visible and near-infrared. The
second project proposes to observe the Sun at mid-infrared and sub-millimetre wavelengths to investigate the physical mechanisms responsible for solar coronal mass ejections. The
final project aims to identify accurate orbits for most of the small-scale Low Earth
Orbiting space debris.


\section{Stellar Properties and Populations}
\subsection{Stellar Populations in Local\\Group Galaxies}

Here, we present a science case for deep imaging of stellar populations with PILOT over degree-sized fields-of-view with an effective ``seeing" substantially better than the best sites today.

These capabilities are essential to tracing stars in crowded regions, and in the outer
parts of galaxies to separate stars from background galaxies. In recent years, stellar
populations in nearby galaxies have begun to reveal tantalizing clues to their formation. Continued success in this field demands that we reach magnitude limits comparable to the
Hubble Space Telescope but over much larger fields-of-view. It is essential to observe at a range of distances, from \si1 out to at least 10~Mpc, in order to cover a sufficient range in local density contrast with respect to the cold dark matter background, with still greater gains if we can reach the Virgo cluster. Only then can we relate the richness of stellar populations and star formation history to the environment.

An optical and near-infrared survey of Local Group galaxies with PILOT at a level of detail far beyond the reach of current theoretical simulations
will provide an observational view of the basic galaxy formation physics that can complement and provide input into simulations of galaxy formation at high redshift. Such a survey will also provide fundamental insight on the process of local galaxy formation. Will we find the edge or will the stellar disc extend even beyond the HI disc? Just how far do baryons extend within the dark matter halo? How extended are bulges in galaxies beyond the Local Group, and do bulge properties have influence on the discs?  There are no reliable predictions to date.

\subsubsection{Discs, Bulges, and Halos in Nearby Galaxies}

On ongoing fundamental question in astrophysics is: how do disc galaxies form and how do
they evolve through cosmic time? There have been important new observations concerning
discs and spheroids in recent years showing that there is still a great deal to learn.
Recent baryon inventories, for example, suggest that most stars are in spheroids today \citep[e.g.,][]{Fukugita_e_98}, while proper bulge/disc deconvolution reveals that most stars are in now in discs and, furthermore, that discs account for a large fraction of the red light in galaxies \citep{Benson_e_07,Driver_e_07}. Discs are also the most fragile of galaxy components, and hence the most sensitive to the physics of galaxy formation \citep{Robertson_e_06}. Therefore, discs lie at the forefront of galaxy formation and evolution studies.

Resolved stellar populations have now been studied in four galaxies (the Milky Way, M31, M33, and NGC 300). Three of these discs (the Milky Way, M31, NGC 300) show no evidence of truncation out to 10 optical scale lengths or more \citep{Worthey_e_05,Yong_e_06,Vlajic_e_08}. \citet{Pohlen_e_08} have shown that outer
discs appear to come in three flavours: endlessly exponential, truncated or flattening
off. These observations were established with surface photometry and remain to be
verified with star count observations of a large survey sample.

Bulges are intimately linked to discs. Bulges are thought to primarily form either from
early monolithic collapse \citep{Eggen_e_62}, through major mergers
\citep{Kauffman_e_94}, or through disc instabilities during the rapid early build-up of
the discs \citep{Bower_e_06}. Bulges provide the link between disc and elliptical galaxies, as they share similar surface brightness distributions and predominantly red colours with ellipticals, indicating their dominance by old stellar populations. Thus, understanding the properties of bulges in comparison and contrast to discs and elliptical galaxies is a crucial step towards a unified picture of galaxy formation.

Substructure in galaxy haloes is a topic of enormous current interest. A remarkable
result was the discovery of the Sgr stream \citep{Ibata_e_94}, which is now known to to accounts for about half the stellar mass of the Galactic halo. Deep observations of M31 and other nearby discs have revealed spectacular streams in the outer haloes
\citep[e.g.,][]{Ferguson_e_02}, thought to represent the satellite accretion histories of massive galaxies \citep[e.g.,][]{Bullock_Johnston_05,Helmi_08}. Whether satellite accretion is responsible for nearly all of halo formation, and the degree to which satellite accretion contributes to disc assembly, remains unknown.

\begin{figure*}[t!]
\begin{center}
\includegraphics[width=7.5cm]{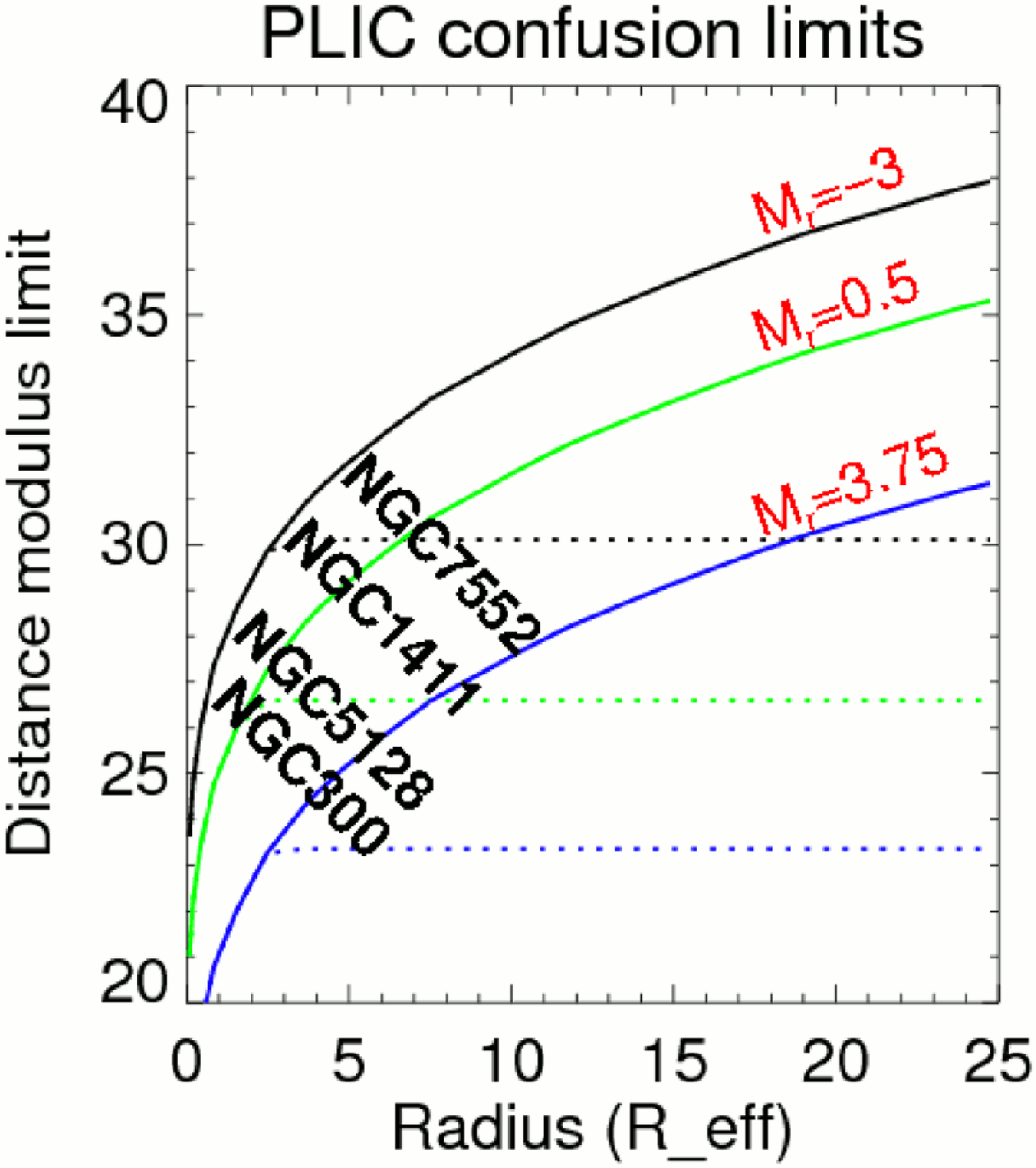}
\includegraphics[width=7.5cm]{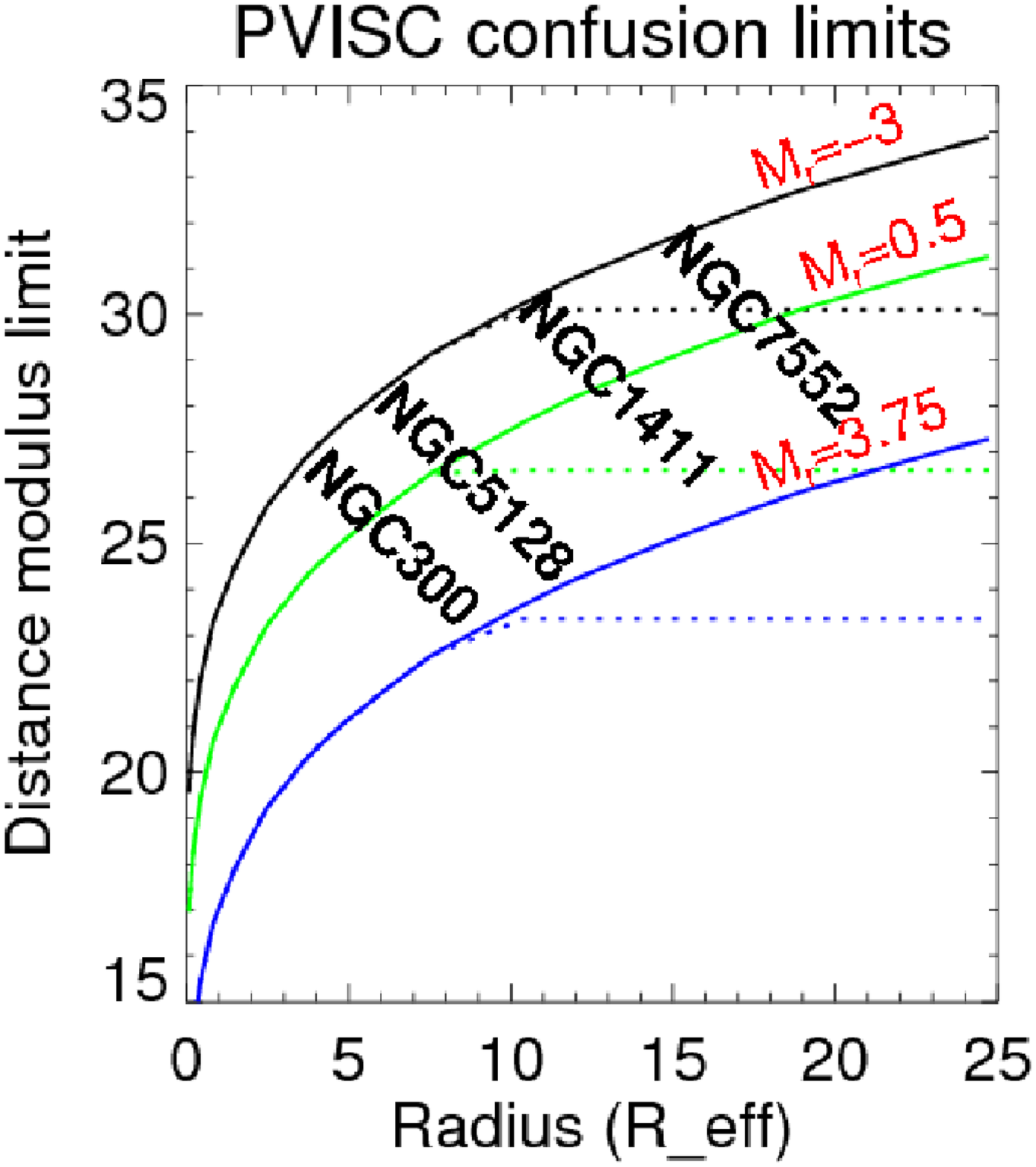}
\includegraphics[width=7.5cm]{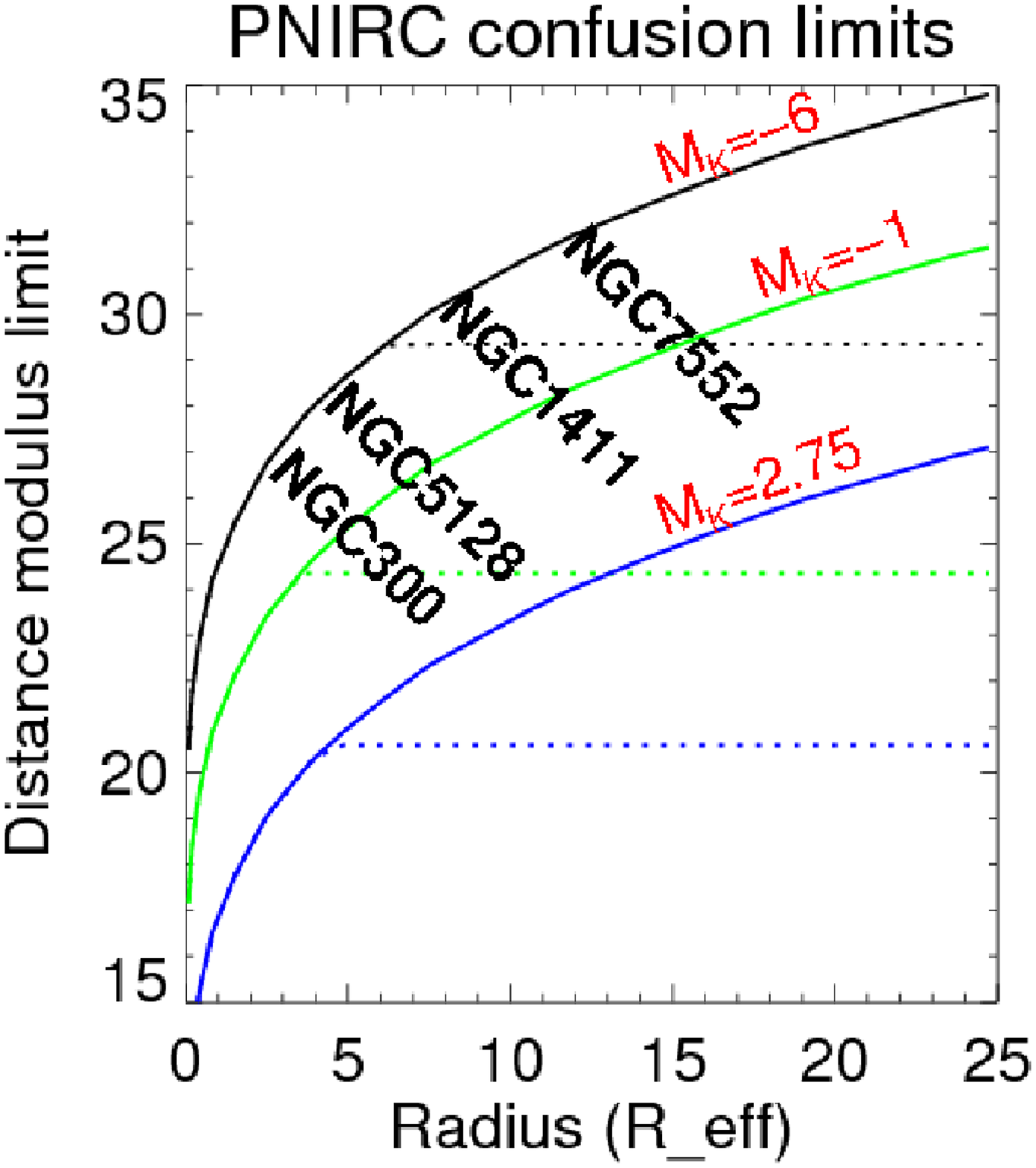}
\caption{Confusion depths for PILOT with PLIC, PNIRC, and PVISC cameras. In these figures, we explore the ability of PILOT to resolve stars with selected intrinsic luminosities, assuming that the target galaxies have surface brightness profiles like that found by \citet{Irwin_e_05} and \citet{Ibata_e_07} along the minor axis of M31 (with $R_{eff} = 0.1\degs$). The solid curves show the distances at which photometric error due to confusion would reach 10\%, as a function of radius along the assumed surface brightness profile. Individual curves are shown for RGB tip stars (black lines), horizontal branch stars (green lines), and main sequence turnoff stars (blue lines), assuming an age of 10~Gyr and metal abundance $Z=0.0001$ and the \citet{Girardi_e_02} isochrones. The confusion limits were calculated following \citet{Olsen_e_03}, by relating the photometric error induced by confusion to the telescope resolution, the magnitude of the target star, and the distance, surface brightness, and luminosity function of the star's environment.  Dotted lines indicate the distances at which photon noise error in a 1-hour exposure would exceed the error due to crowding.}\label{local_confusion}
\end{center}
\end{figure*}

\subsubsection{Sensitivity and Crowding}

Stellar photometry depends critically on an accurate measure of the background,
fluctuations in which impose a fundamental limit on the possible accuracy of the
photometry. In Figure~\ref{local_confusion}, the crowding limits for various target
galaxies observed with the PLIC, PVISC, and PNIRC cameras of PILOT are estimated
following the analysis of \citet{Olsen_e_03}. The crowding limit is defined to be the
magnitude at which the photometric error due to crowding is 0.1~magnitudes. This limit depends on the telescope resolution, the magnitude of the target star, and the distance, surface brightness, and luminosity function of the star's environment.  For the purposes of this investigation, we assumed that all target galaxies have surface brightness profiles in their outer disks like that found by \citet{Irwin_e_05} and \citet{Ibata_e_07} along the minor axis of M31, who measured surface brightnesses between 25 and 29 mags\,arcsec$^{-2}$ at \vb\ at radii between 8 and 18~kpc from the center.

With the high resolution Lucky Imaging camera, PLIC, PILOT will achieve photometry in
nearby galaxies as deep as with HST/ACS. In the outskirts of the Magellanic Clouds, PLIC
will reach the bottom of the stellar luminosity function, whereas in Sculptor group
galaxies, such as NGC 300, the photometry will reach well below the level of the
horizontal branch. The confusion limits for PVISC are similar to those of PLIC at much
lower surface brightnesses, reaching the level of the horizontal branch out to
\si3.5~Mpc. At distances of \si7~Mpc, PVISC will still resolve stars down to below the
red giant branch tip. PILOT+PVISC, with its large field and excellent resolution, will
thus be a powerful instrument for mapping the properties of extended galaxy discs and
halos. Observing in the near-infrared with PNIRC makes it possible to resolve red giants
out to \si10~Mpc, and AGB stars out to the Virgo cluster, owing to the very high
luminosities of these stars in the near-infrared.

\subsubsection{Observational Approach}

A program of deep, wide-field imaging of a sample of galaxies out to \si10~Mpc would
allow us to answer a number of very exciting questions. We would aim to understand the
relationship between extended discs, inner discs, bulges, and halos, and to control for
the cosmologically predicted effect of the density of the environment.

The program is challenging for many reasons, however. First, we need to reach a
photometric depth sufficient to determine metallicities from stars a few magnitudes down
the metal-poor red giant branch; to push beyond the Local Group, we must reach a
(5--10~$\sigma$) point source sensitivity to $\mab\approx28$ in \emph{g}- and
\emph{r}-bands, and $\mab\approx27$ in the near-infrared at \kb. It must be stressed that accurate stellar photometry below this limit in any band is very challenging. Excellent
seeing, stable photometric conditions, and a dark site are essential, as are high
instrument throughput and a stable flatfield. Second, the need for global coverage and
the extents of the target galaxies requires wide fields in order to map several square
degrees per galaxy. Finally, for galaxies at these distances, crowding by point sources
is a serious limiting factor under conventional seeing conditions, even in the low
surface brightness extended discs.

The goal of this study is to observe a sample of disc galaxies over a range of
environmental density. We will use PLIC to study the high surface brightness inner parts
of the discs, and PVISC and PNIRC to map the low surface brightness extended discs. PVISC observations reaching down to the horizontal branch in galaxies within \si3.5~Mpc (and
thus chiefly low density environments) will allow us to study the detailed star formation histories of extended discs, while we will use PVISC and PNIRC observations of red giant
and AGB stars in more distant galaxies to map the prevalence and extent of such discs in
many different environments. We will primarily target moderately inclined and face-on
discs at reasonably high Galactic latitude, as these avoid projection effects in our
analysis of population gradients and minimize Milky Way foreground extinction. We also
need to observe as many galaxies as possible within each environment, to control for the
variance in the interaction histories of the galaxies.

The actual magnitude limit is a shallow function of distance, as the idea is to reach the limit imposed by confusion in each galaxy. For an accurate metallicity determination and star/galaxy discrimination, 3 or 4 filters, likely to be \emph{g}, \emph{r}, \emph{J},
\kd, will be required for each galaxy; the exact choice will depend on the particular galaxy to be studied.  To detect the TRGB out to 10~Mpc in emph{g} and \emph{r} will require \si1 hour with PVISC, and in \emph{J} and \emph{K} will require \si3.5 hours with PNIRC.  Detecting the horizontal branch out to 4~Mpc will require \si10 hours of exposure with PVISC in \emph{g} and \emph{r}, and \si1 hour with PNRIC in \emph{J} and \emph{K} to get \si2 magnitudes below the TRGB.  Our goal is to sample the populations in the disks out to $10-15 R_{eff}$, the radius out to which M31's disk has been traced \citep{Irwin_e_05,Ibata_e_07}. For galaxies with distances $1.5 < d < 10$~Mpc, we can capture the necessary area with one PVISC pointing, whereas with PNIRC we would use \si3 pointings per galaxy to probe the range of radii. Guided by the \citet{Tully_87} catalog of nearby galaxies, we find that PILOT will have 22 possible targets for TRGB measurement and 10 targets for horizontal branch measurement. Given the rough exposure times and number of pointings above, we could observe the entire sample of 32 galaxies in \emph{g}, \emph{r}, \emph{J},
\kd, in \si800 hours during one complete winter season.

The science case outlined here demands a telescope with excellent resolution for overcoming stellar crowding, and a wide field for mapping and characterizing extended stellar discs in galaxies over a range of environments. The Antarctic site conditions, and the PILOT telescope in particular, provide a very good match to these requirements. These studies are important for understanding how disc galaxies formed, as they track the build-up of stellar mass, and place limits on the merger histories of disc galaxies. They do not, however, have the resolution to study high surface brightness discs beyond distances of \si4~Mpc, and thus do not reach the richest galaxy environments. The recent discovery of extremely low surface brightness extended stellar discs is thus very exciting. Not only do they offer a fresh perspective on the formation history of disc galaxies generally, but they give us the opportunity to trace stellar disc formation out to larger distances and thus richer environments, where we expect to be able to see differences in the hierarchical build-up of galaxies compared to those in the Local Group.

\subsection{Outer Structures of Satellite\\ Galaxies}
This project aims to observe the stellar content of the outer regions of nearby Milky Way satellite galaxies (out to \si600~kpc) that are sufficiently nearby to allow for their
individual members to be resolved. Galaxies with a resolvable stellar content provide a
useful tool to understand the process of galaxy formation and evolution. Tracing the
distribution of stars of a different type, age, and metallicity allows us to access the
propagation of the star formation history throughout galaxies and the morphology of
galaxies as a function of time.

The outermost parts of galaxies harbour faint and old stars that, according to
cosmological simulations, may be associated with a galaxy halo and bear the signature of
tidal streams resulting from galaxy interaction and satellite accretion events. Even the
halos of those galaxies that did not experience major mergers preserve the fossils of
their formation history which can be revealed by wide-field sensitive studies of halo
objects such as red giant and horizontal branch stars. On the contrary, galaxy halos may
be rich in substructures like stellar clusters. One of the challenges of modern astronomy
is to understand if stellar clusters with multiple stellar populations are the relics of
dwarf galaxies that were accreted at earlier times.

\begin{figure*}[t]
\begin{center}
\includegraphics[width=12cm]{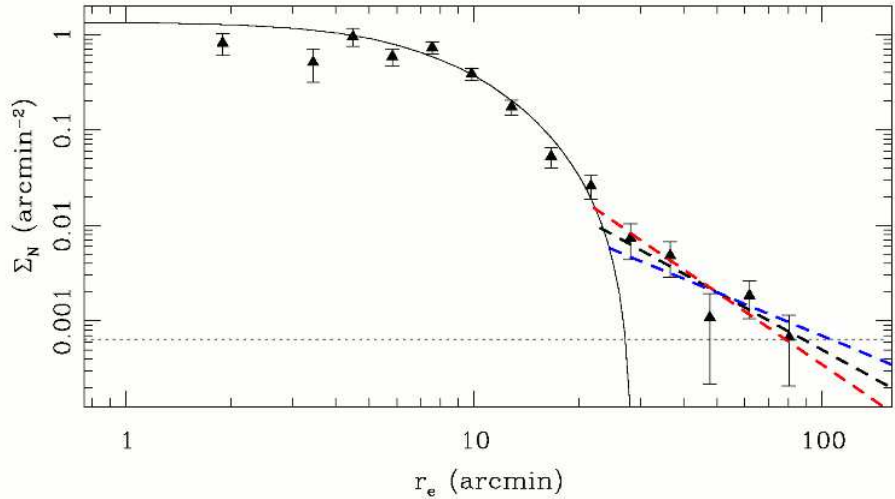}
\caption{Density profile for the Carina dSph galaxy showing extended stellar components
at large radii. From \citet{Munoz_e_06}. PILOT will enable the stellar content in these
outer regions of such galaxies to be analysed in more detail than currently possible.
}\label{satellite_profile}
\end{center}
\end{figure*}

This project aims to observe the outer regions of galaxies that are sufficiently nearby
to allow for individual field and cluster members to be resolved. The main goal is
tracing the outer morphology, structure, age, and metallicity of galaxies. This will have
a direct impact on our understanding of the process of galaxy formation and evolution. It
will complement current studies on the most crowded, active and central regions of
galaxies \citep[e.g.,][]{Holtzman_e_00} addressing, for example, issues of star
formation. It also provides a complement to the Local Group galaxy survey proposed in the
previous section, as it will provide greater detail on the outer structures of closer
galaxies.

The outer regions surrounding nearby galaxies have been mostly studied in a limited
number of fields \citep[e.g.,][]{Demers_e_06,Gullieuszik_e_07}. These investigations
successfully probed the presence of an old stellar population which is much more extended
than initially expected \citep{Majewski_e_05,Munoz_e_06}, as illustrated in
Figure~\ref{satellite_profile}. At present, and for many more years to come, VISTA will
be the only type of facility that can be used for these studies. Its spatial resolution
(\si0.34$''$/pixel for a natural seeing of at best \si0.6$''$) is the only limitation
restricting the galaxies that this study can probe out to a given distance. The small
field-of-view of planned infrared instruments for ELTs and JWST precludes their use for
wide-area survey projects.

The major contribution of PILOT to this research will be the combination of high spatial
resolution and wide-field coverage at near-infrared wavelengths. The possibility of
exploiting a natural seeing that is considerably better than that available for a similar
wide-field instrument at other ground-based site will allow us to target galaxies that
are further away and, most importantly, to distinguish fainter sources in their
outskirts.

The old and extended stellar population of galaxies is traced either by red giant branch
(RGB) stars or by RR Lyrae variable stars. RGB stars are usually more numerous than RR
Lyrae stars and from the bottom of the RGB to the tip of the RGB they span \si8 mag in
the \kb. RR Lyrae stars have a luminosity which is about 6 magnitudes fainter than the
tip of the RGB.

RR Lyrae stars are easily identified in optical wavebands from their light-curve. In the
\kb\ their amplitude of variation is reduced and with just a limited number of epochs it
is possible to obtain a very good estimate of their mean magnitude. The period-luminosity
relation for RR Lyrae stars in the \kb\ has a small scatter and depends very little on
age and metallicity providing a very powerful distance indicator
\citep{Pietrzynski_e_08}. The combined number of epochs that will be used for the study
of RR Lyrae stars will also be sufficient to detect most RGB stars.

The 1 hour sensitivity limit with PILOT is $\mab = 25.3$ at \kd\ for 5~$\sigma$ (which is
the minimum SNR required for this study). This would allow the detection of the entire
RGB and RR Lyrae stars in galaxies out to distances of \si600~kpc where we find most of
the satellites of the Milky Way. The Magellanic Clouds are being extensively studied with
other facilities such as VISTA in the \emph{Y}, \emph{J}, and $K_s$ wavebands via the VMC
survey; a deeper and higher wavelength study of the Magellanic Clouds is proposed for
PILOT in the next Section. The remaining Milky Way satellites span a range of
morphological types and histories that are perfectly suitable for this study. Targets
within 600~kpc include Sculptor, Carina, Fornax, Phoenix, and NGC~6822. The main body of
these galaxies covers an angular scale ranging from $5'\times4'$ to $40'\times31'$.

To survey each of these galaxies out to three times the extent of the galaxy in both
directions will require a total coverage of \si5~\degsq. This equates to \si180~fields
with PNIRC. Multi-band observations in at least $Y$, $J$, and \kd\ filters, and multiple
observations of each field at different times are required. The indicative total
observing time for this project is \si4000~hours (i.e., \si2~winter seasons).

\subsection{A Magellanic Clouds Broadband Survey}

The Magellanic Clouds (MCs) offer a unique opportunity to investigate in great
detail---at the level of individual stars---the history of the evolution of a complex
system of gas and stars and the recycling of matter within a relatively limited area of
the sky (about 100~\degsq) where the distance is accurately known.

The programme proposed here for PILOT is to map fully and uniformly the surface of both MCs and selected areas of the Magellanic Stream at an unprecedented level of sensitivity and angular resolution in the \kd\ and \ls, and possibly the \emph{L}$'$ and \emph{M}$'$, bands. This will complement and considerably enhance the database already obtained with 2MASS/DENIS and the InfraRed Survey Facility (IRSF) in the non-thermal infrared (up to \ks), and Spitzer (SAGE) in the mid- and far-infrared, or planned to be obtained with VISTA (VMC) up to \ks. This survey will basically investigate:
\begin{enumerate}
\item The star formation processes in an environment of lower metal abundance than in our Galaxy and subject to strong tidal effect from our Galaxy, and

\item Extreme populations of AGB stars characterized by a high mass loss rate and
    thick circumstellar envelopes that may have escaped detection by earlier deep
    2~\mm\ surveys or not resolved enough or confusion limited in the SAGE survey.
\end{enumerate}

This project relies on a number of the advantages that the Dome~C location offers with
respect to conventional sites. It requires the exceptional observing conditions in the
near thermal infrared (improved transmission, low sky emissivity by a factor of up to 3
magnitudes at \kd, better photometric stability, and the opportunity to open new windows
such as the \ls-band at 3--3.5~\mm) and the excellent seeing conditions (0.3$''$) above
the turbulent boundary layer. Additionally, observing efficiency is increased thanks to
the very low background emission of the sky throughout the year; observations may even be possible in summer time, at least in the \ls\ filter and possibly fractions of the summer ``days" in the \kd\ window. Finally, the high latitude of Dome~C means that the MCs are
circumpolar and thus can be observed all year round at small and almost constant airmass
(1.0 to 1.15).

\begin{figure}[t!]
\begin{center}
\includegraphics[width=7.5cm]{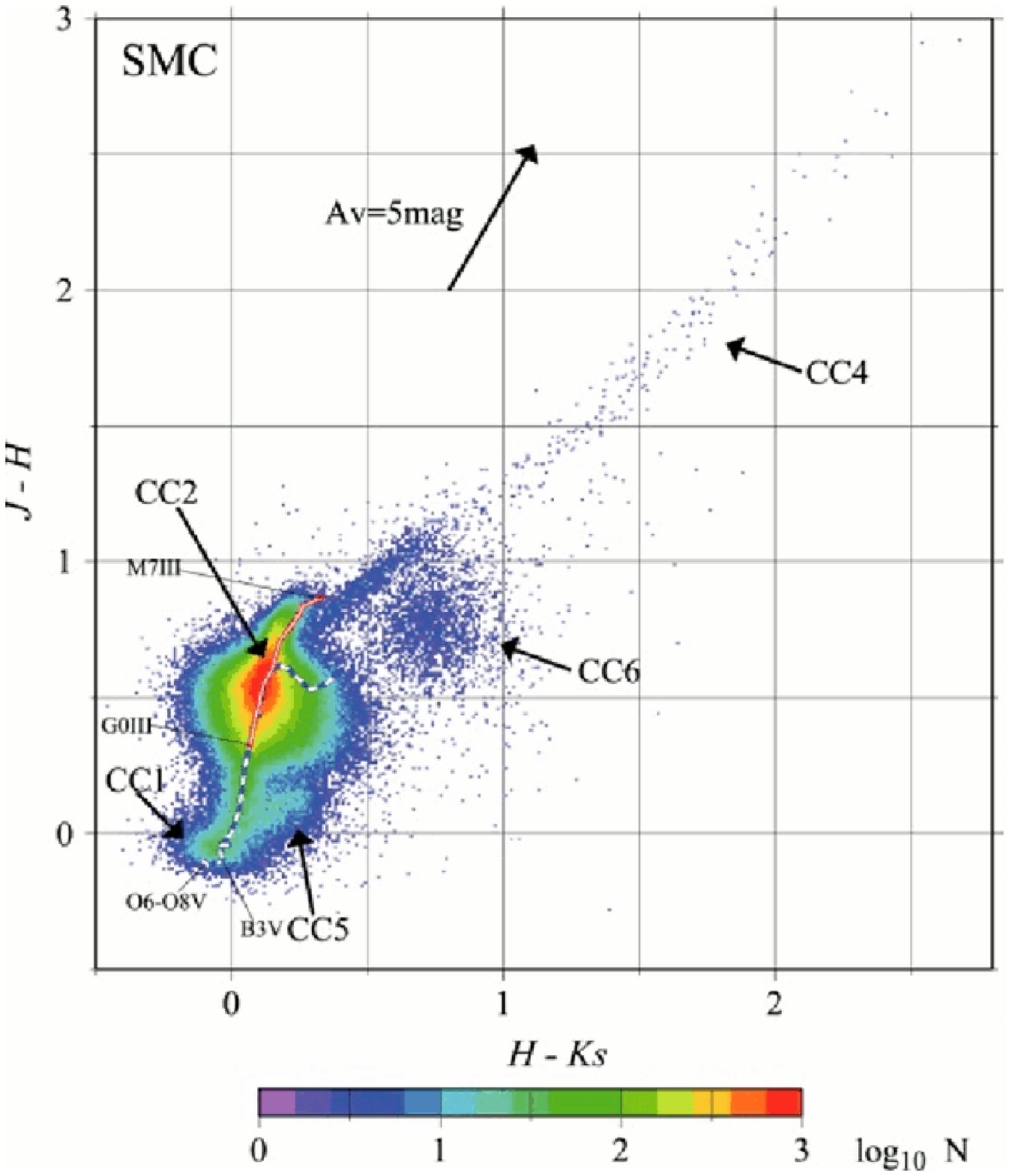}
\includegraphics[width=7.5cm]{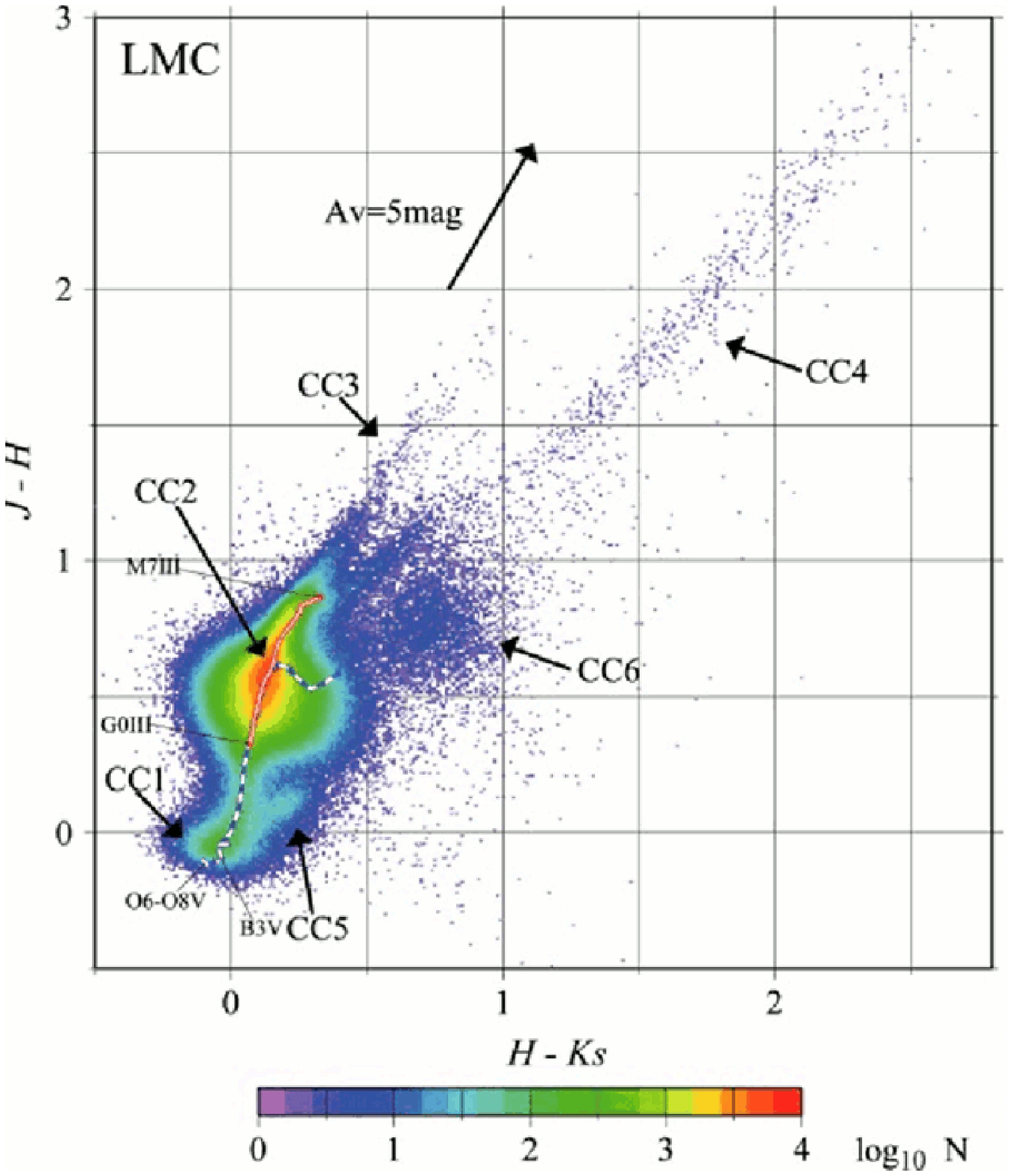}
\caption{Infrared colour-colour diagram of the Large Magellanic Clouds (top) and the
Small Magellanic Clouds (bottom) from the IRSF survey. From \citet{Kato_e_07}. The MC survey proposed here for PILOT would extend to longer wavelengths and deeper
limits, allowing star formation to be probed to lower masses in more obscured
environments. }\label{maclas_IRSF}
\end{center}
\end{figure}

This Magellanic Clouds study is complementary to the 2MASS/DENIS, IRSF and future VISTA Surveys. In the last decade, several near-infrared surveys of the Magellanic Clouds have been performed providing catalogues of millions of stellar entries, in the \emph{I}, \emph{J}, \emph{H}, and \emph{K} bands. These have included DENIS \citep{Cioni_e_00}, 2MASS \citep{Nikolaiev_e_00}, and IRSF \citep{Kato_e_07}. Colour-colour diagrams (see for example Figure~\ref{maclas_IRSF}) have been extensively used to separate different
populations of stars and objects and, thanks to the fact that the objects are essentially
at the same accurately known distance, colour-magnitude diagrams have been used as well,
to provide excellent luminosity calibrations of these populations.

The SPIREX telescope at the South Pole indicated the kind of study that would be
possible, through deep thermal infrared imaging of the 30 Doradus region of the LMC
\citep{Maercker_Burton_05}.

In addition, the Spitzer Legacy program, Surveying the Agents of a Galaxy's Evolution
(SAGE)\footnote{See \url{http://sage.stsci.edu/project.php}}, has recently provided
unique images and catalogues of sources of the MCs \citep[e.g.,][]{Whitney_e_08} in the
mid- and far-infrared. SAGE has undertaken a comprehensive picture of the current star
formation activity, which is traced by the IRAC (3.5, 4.5, 5.8, and 8.0~\mm) and MIPS
(24, 70, and 160~\mm) bands, which have spatial resolutions of 2$''$ and 6--40$''$
respectively.

Starting in 2008, one of the VISTA Public Surveys, VMC\footnote{
\url{http://www.eso.org/sci/observing/policies/PublicSurveys/sciencePublicSurveys.html}},
will be aimed at deep imaging of the MCs and Bridge. This will provide unprecedented
catalogues of faint objects in the \emph{Y}, \emph{J}, and \emph{K} bands to $\mab =
22.9$, 22.8, and 22.3 respectively at 10~$\sigma$ across some 200~\degsq, at a
seeing-limited FWHM resolution of \si0.8$''$. The total number of nights required to
achieve VMC in 5~years is approximately 200. It will resolve stellar populations into old populations, active star formation areas, and ongoing merging. It will globally resolve
the star history with unprecedented accuracy and trace past interactions.

The aim of this survey is to cover in a first campaign the $7\degs\times7\degs$ area of the LMC at \kd\ and \ls\ (with a possible extension to \emph{L}$'$ and \emph{M}$'$) at the sensitivity limit expected in one hour with the PILOT Near InfraRed Camera (PNIRC), i.e., $\mab = 25.3$ and 21.2, respectively, at a spatial resolution of 0.3--0.4$''$. This is \si15 times deeper than VISTA at 2.2~\mm, with 2.5 times the angular resolution, and half as deep as SAGE (5.1~\mj) at 3.6~\mm, but at an angular resolution 5--6 times higher.

This extension will be extremely useful to single out more deeply embedded objects
(younger or experiencing even more intense mass loss) that only show up in the 3--5~\mm\
range, as well as to pick up objects of lower surface temperatures (e.g., extreme carbon
stars). The PILOT MC survey will offer an opportunity to link the VISTA and Spitzer SAGE
data, and will prepare for more detailed observations with JWST and/or a southern ELT.

The baseline near-infrared camera for PILOT, PNIRC, is equipped with a single channel
near-infrared $4\mathrm{k} \times 4$k array covering a $10'\times10'$ field at 0.15$''$
per pixel. This instrument would require about 2000 hours of observation to
cover the 50~\degsq\ area of the LMC at the ultimate depth ($\mab\approx25$ at \kb). This could be achieved in approximately 3~months of continuous and essentially uninterrupted
observations taking into account calibration and overheads. Another 1000~hours would be
necessary to cover the SMC and some selected fields in the Magellanic Stream at the same
level of sensitivity.

The efficiency of this survey is a strong function of the instrument specification. The
efficiency would be greatly increased by enlarging the focal plane coverage, either by
using larger arrays (i.e., with a mosaic of $4\mathrm{k} \times 4$k arrays), or a
multi-channel configuration (e.g., with \kd, \ls, and \emph{L}$'$ arms). The latter
option has the advantage of providing simultaneous photometric data, with no phase lag
for variable objects, and thus accurate colour indices. The ideal configuration involves
a trade off between cost, observing lapse of time, and requested sensitivity. The most
important issue is probably to reach the best gain in sensitivity compared to the VISTA
survey in the \kb.

\subsection{Asteroseismology of stars in clusters }

The study of stellar oscillations, known as \emph{asteroseismology}, allows us to probe
the interiors of stars. Oscillation periods range from a few minutes up to days or even
years, and the photometric amplitudes range from a few parts per million up to a
magnitude or more. Asteroseismology is a rapidly growing field that covers a wide range
of stars, including Sun-like stars, red giants, hot massive stars and white dwarfs.

Recent results have demonstrated the potential for time-series photometry from Dome~C
\citep{Rauer_e_08,Strassmeier_e_08}. With PILOT, we will be able to apply the techniques
of asteroseismology to stars in clusters. The projects described below involve observing
oscillating stars in two types of clusters using the PILOT Visible Camera (PVISC), taking
advantage of the following capabilities:
\begin{itemize}
\item A large field-of-view with strongly reduced crowding problems thanks to the
    exceptional seeing.  Only HST can to do better in resolution, but it has a much
    smaller field.
\item Nearly-continuous temporal coverage, which means improved precision in
    frequency determination, without the ambiguities associated with
    one-cycle-per-day aliases.
\item A much smaller range of airmasses than at low-latitude sites, since a given
    field stays at roughly constant elevation as the Earth rotates.  This should
    yield photometry whose precision on time-scales of hours to days is exquisite.
\item Photometric precision will be further improved by the low scintillation noise
    at Dome~C, which should allow us to detect much smaller fluctuations than at any
    other ground-based observatory.
\end{itemize}

The importance of asteroseismology is seen in the number of projects underway or being
planned. The French/ESA CoRoT space mission is currently collecting data and the NASA
Kepler mission was recently launched in March 2009. CoRoT is observing a handful of pre-determined fields in its continuous viewing zones\footnote{See
\url{http://smsc.cnes.fr/COROT/}}, while Kepler will study a single field in Cygnus (with the primary aim of detecting transiting exoplanets\footnote{See \url{http://kepler.nasa.gov/}}). Neither has the capability to observe clusters in the
way proposed here. On the ground, the desire to obtain continuous coverage has led to
plans for global networks of telescopes at temperate sites, to carry out observations in
photometry \citep[LCOGT;][]{Hidas_e_08} and radial velocity
\citep[SONG;][]{Grundahl_e_06}. Again, these are geared towards individual stars and do
not have the capacity to observe clusters. The same applies to the SIAMOIS instrument,
which has been proposed for Dome~C by \citet{Mosser_e_07}. If funded, it will measure
precise stellar velocities and allow asteroseismology of a small number of bright
solar-type stars, complementary to, but very distinct from, the observations of huge
numbers of much fainter cluster stars, to which PILOT is uniquely suited.

\begin{figure*}
\begin{center}
\includegraphics[width=5.2cm]{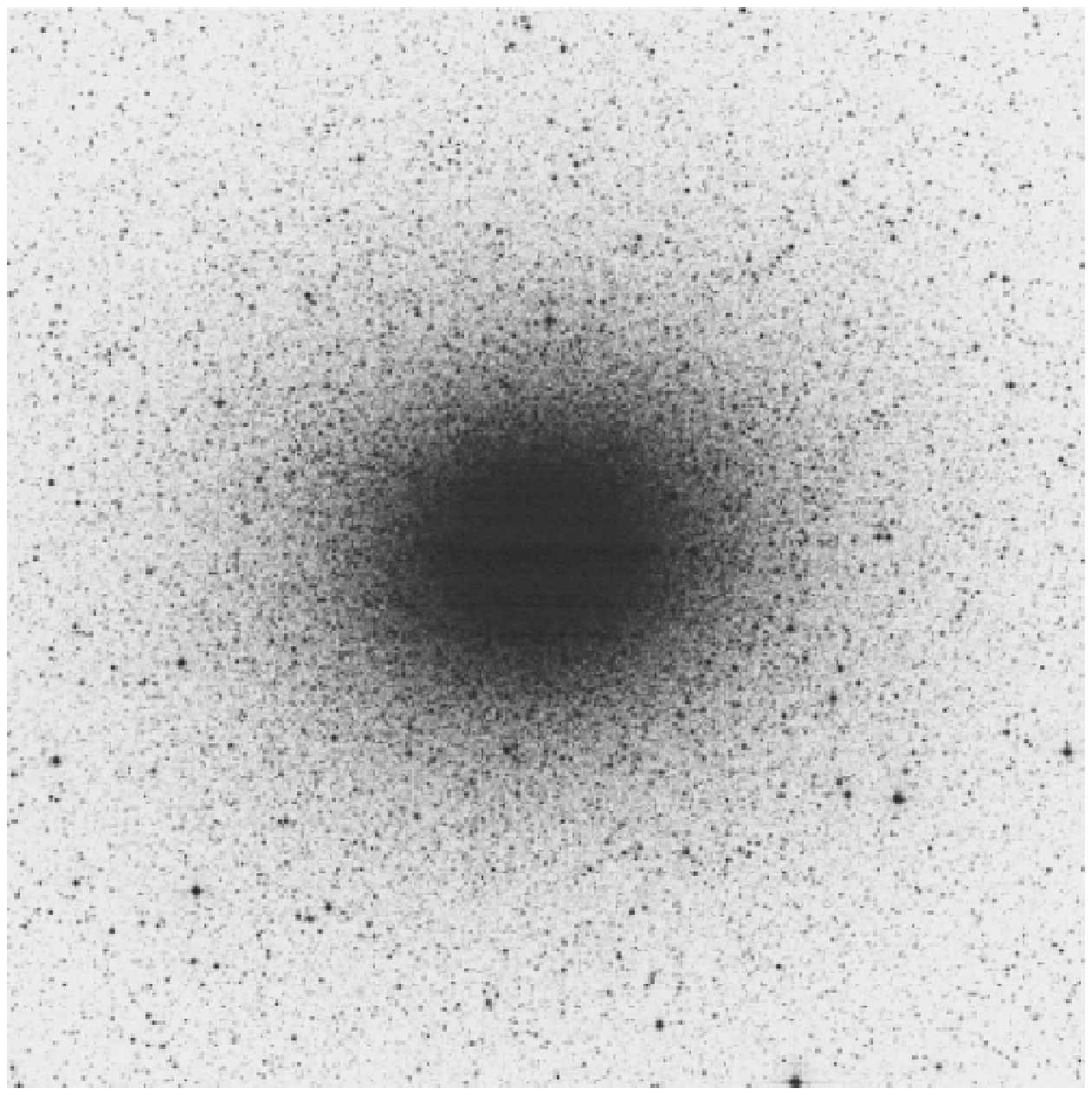}
\includegraphics[width=5.2cm]{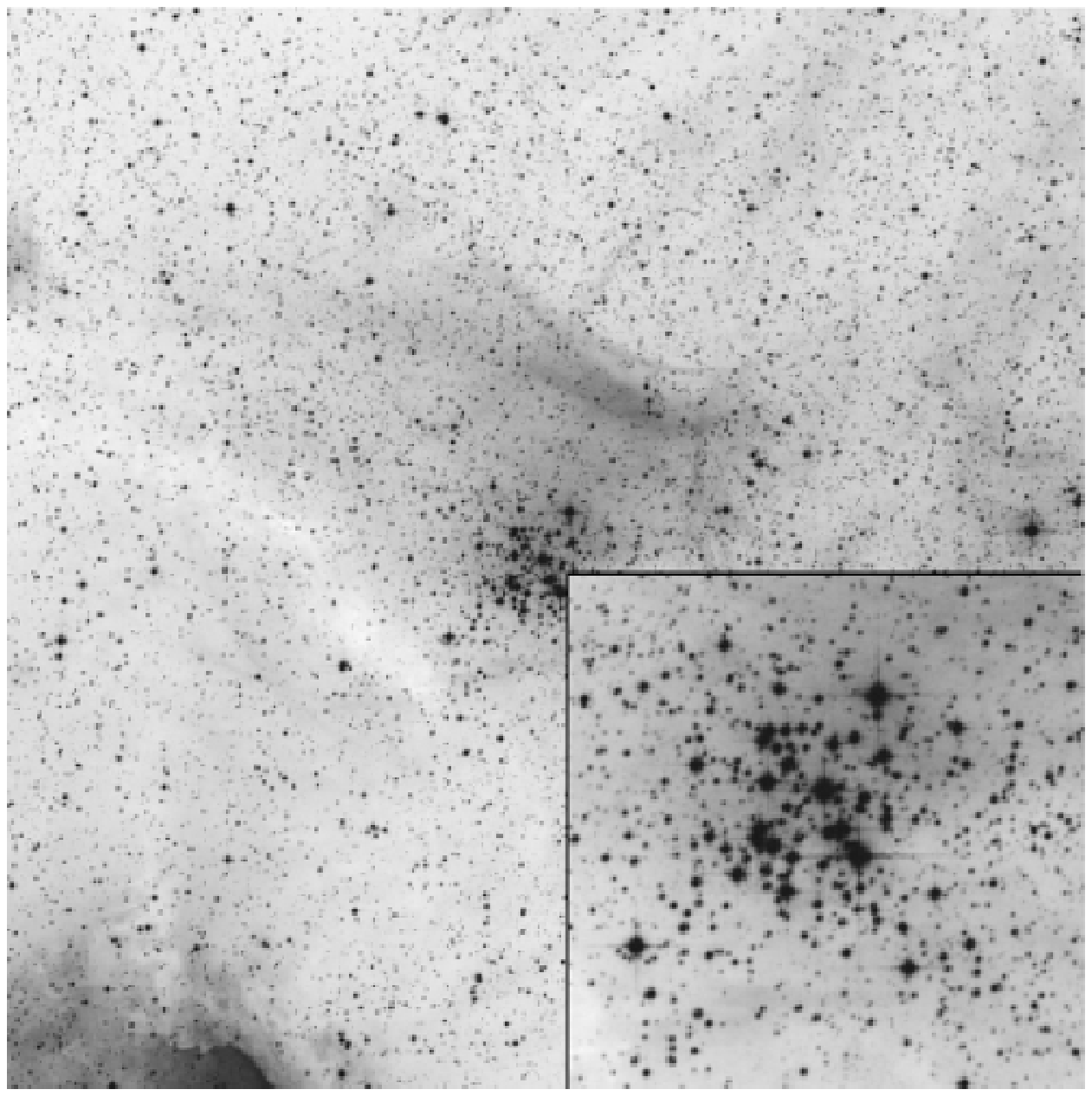}
\includegraphics[width=5.2cm]{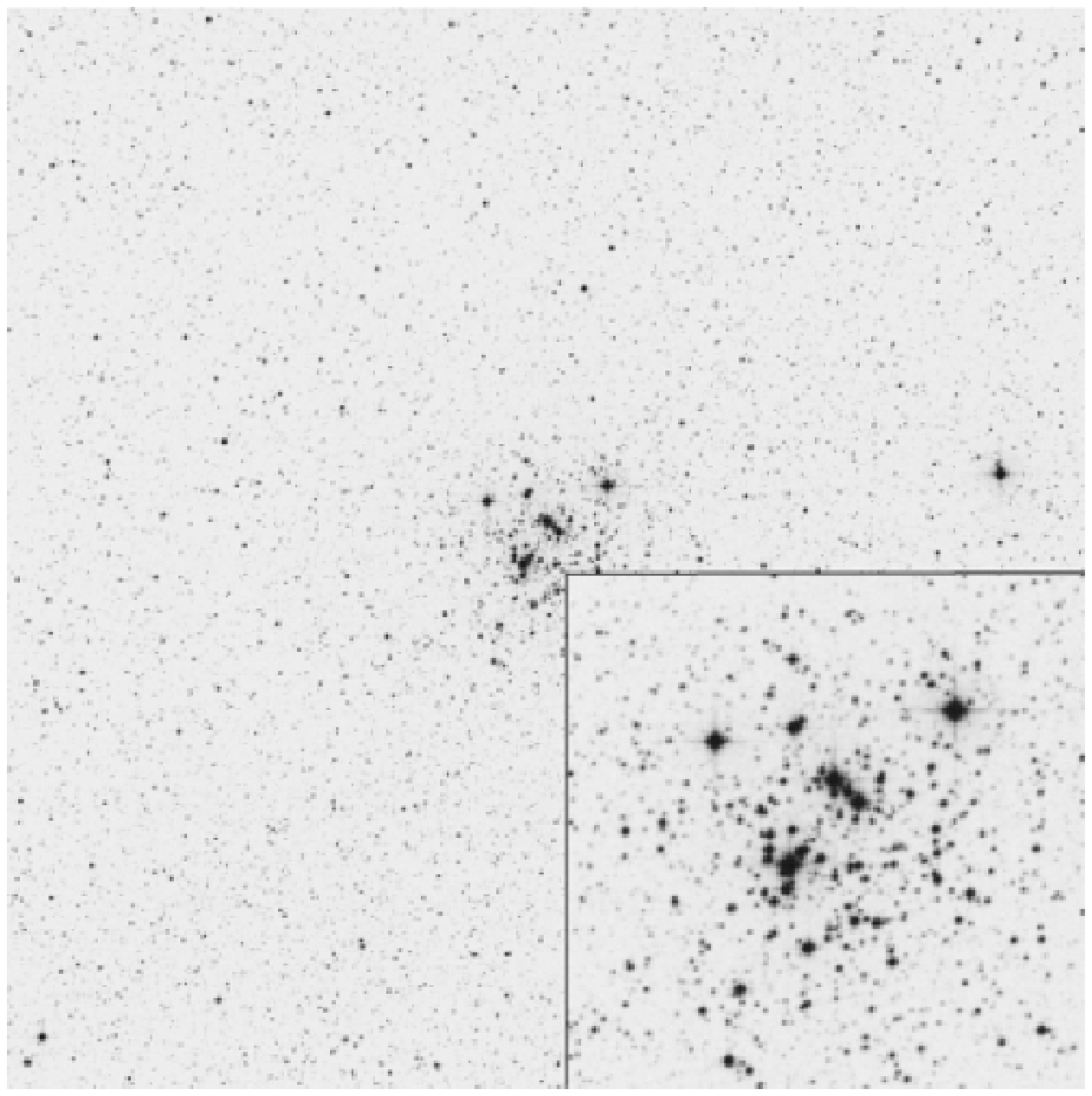}
\caption{The field-of-view of PILOT is perfect for \ocen\ (left panel) and will fully
cover both NGC 3293 (middle) and NGC 4755 (right).  For NGC 3293 and NGC 4755 an expanded
view ($\times$ 2) of the central part of the cluster is shown in the lower right part of
the panel.  The three images are all $42' \times 42'$ (\si the FOV of PVISC) and obtained
from the Digitized Sky Survey. Note that the field-of-view of WFPC3, to be installed on
HST in 2009, is $2.7' \times2.7'$ (0.4\% of the area covered by PILOT).
\label{astro_clusters}}
\end{center}
\end{figure*}

\subsubsection{\ocenn: Probing Star Formation in the Early Universe}

Globular clusters (GCs) contain 10$^5$--10$^6$ very old stars within roughly
spherical aggregates that are several tens of parsecs in diameter (see
Figure~\ref{astro_clusters}). Typical ages range from 10 to 13~Gyr, with strong
chemical and age homogeneity within each cluster. These objects are enigmatic
ingredients in stellar and Galaxy formation, offering an opportunity to unravel
cosmic history through careful observations. In particular, one current understanding of the formation of the Milky Way Galaxy envisages hierarchical merging of low-mass building blocks that were formed in mini-halos at redshifts $z\approx5$--10
\citep[e.g.,][]{Kravtsov_Gnedin_05,Bekki_e_07}. Within this framework the 150 or so
globular clusters in the Milky Way system \citep{Harris_96} are thought to represent
the survivors from these original building blocks.  The small range of metallicities
in individual clusters and the large cluster-to-cluster variations could indicate a
chemical diversity among the mini-halos reflecting the cosmic variance of initial
conditions of star formation.

\begin{figure*}[t!]
\begin{center}
\includegraphics[width=7cm]{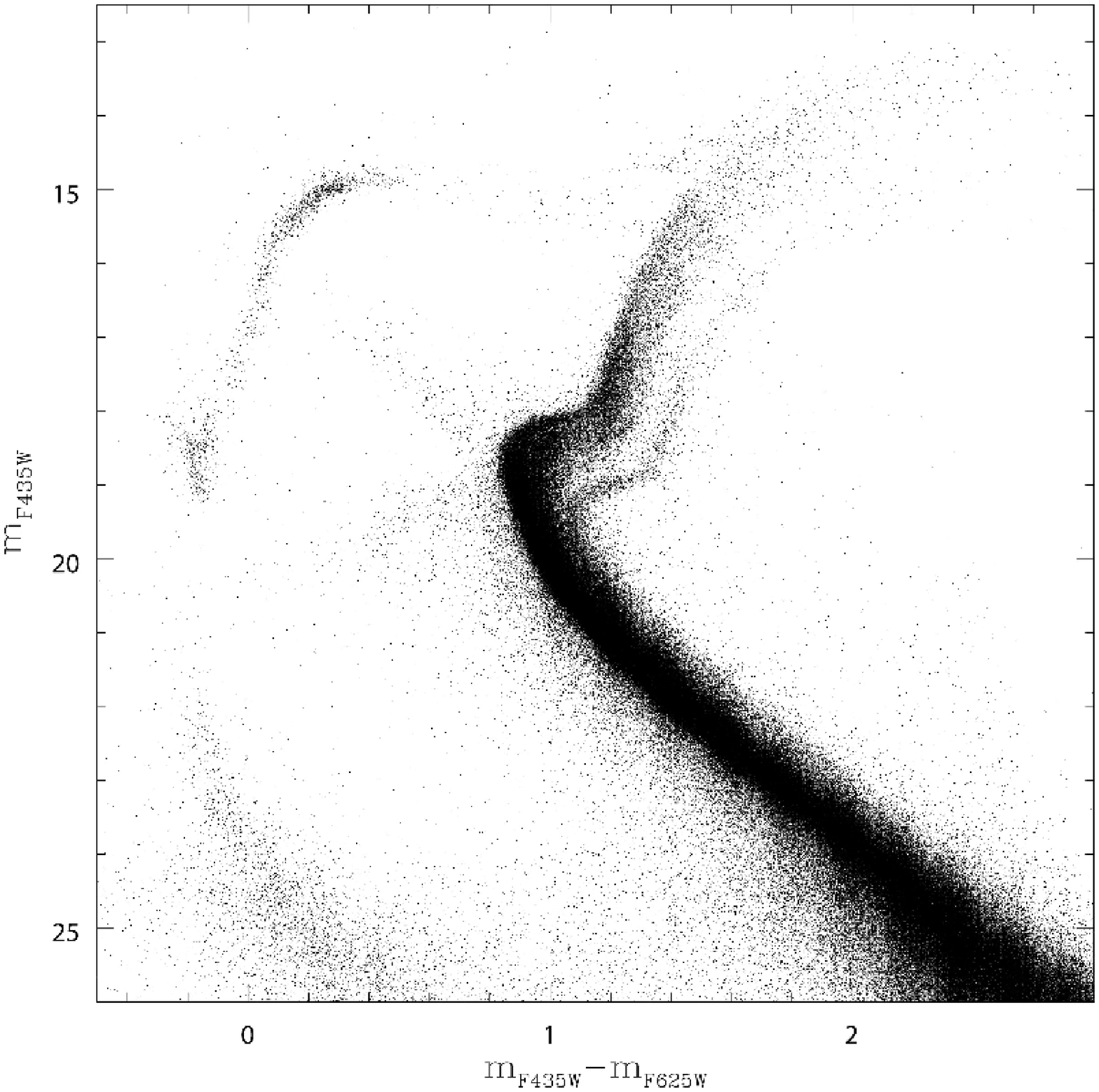}\\
\vspace{0.5cm}
\includegraphics[width=6cm]{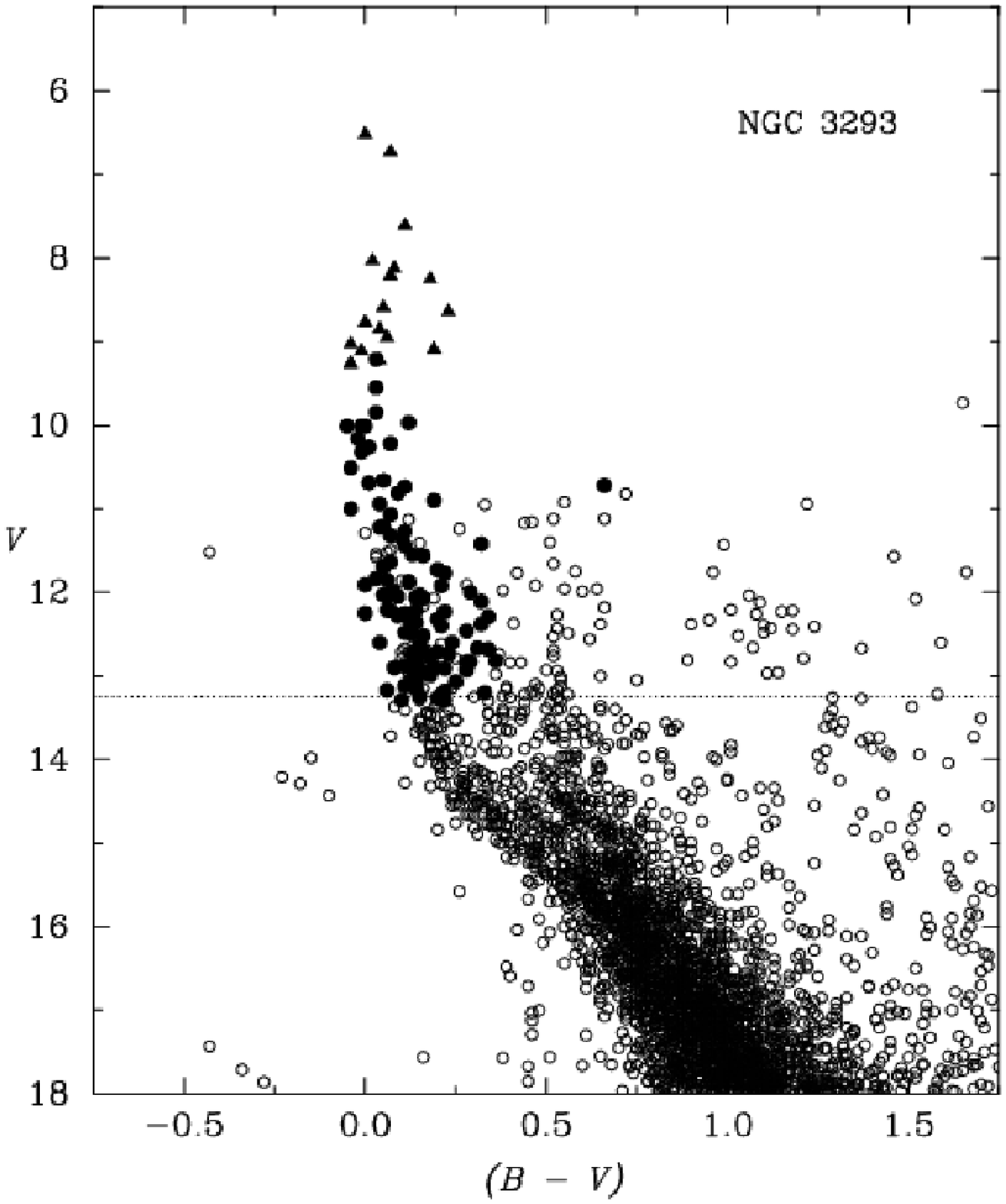}
\hspace{0.5cm}
\includegraphics[width=6cm]{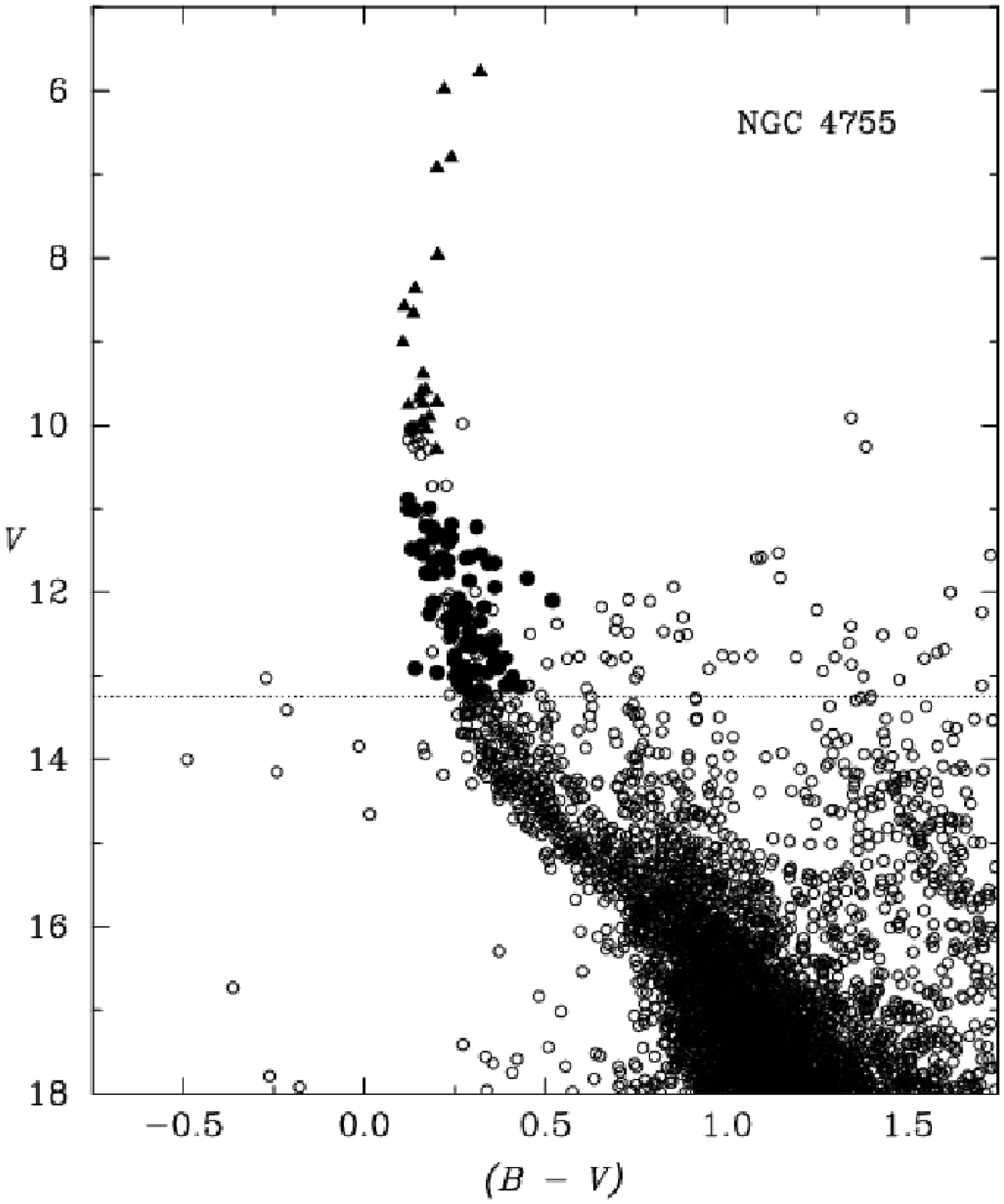}
\caption{Colour-magnitude diagrams of \ocen, NGC 3293 and NGC 4755. Figures from
\citet{Villanova_e_07} and \citet{Evans_e_05}. \ocen\ has an extended horizontal branch
and a large white dwarf population.  The age of \ocen\ is about $12$~Gyr.  The open
clusters are much younger ($\sim10$~Myr) and will allow us to probe B- and A-type stars
on the main sequence. \label{astro_cmd}}
\end{center}
\end{figure*}

The discovery of multiple stellar populations in \ocen, the most massive and
exceptional GC of the Milky Way \citep{Lee_e_99}, together with the fact that the second most massive GC (M54) is a core of the disrupting Sagittarius dwarf galaxy \citep{Layden_Sarajedini_00}, have strengthened the view that some of the most massive GCs might be the remnant cores of disrupted nucleated dwarf galaxies. The peculiar colour-magnitude diagram of \ocen\ (see Figure~\ref{astro_cmd}) contains multiple branches, indicating a mixture of distinct stellar populations within the cluster. Two basic scenarios have been proposed to explain the spread in chemical composition and the other unusual properties of \ocen. The first is self-enrichment from the first-generation asymptotic giant branch stars and Type II supernova events, either as an isolated cluster or as the nucleus of a dwarf galaxy. The second scenario proposes some sort of merger event, such as a merger between two or more globular clusters or between a dwarf galaxy and a globular cluster. There has been a great interest in testing these scenarios and the evidence is rather inconclusive \citep[e.g.,][]{Platais_e_03,Pancino_e_07}.

We propose a new approach to investigate the properties of distinct populations, using
asteroseismology in combination with photometry and spectroscopy to derive
accurate stellar parameters along the red giant and subgiant
branches of \ocen. In particular, asteroseismology will give us constraints
on the stellar age to about 10\% and the mean stellar density to less than
1\% \citep{Kjeldsen_e_09}. We will use the unique capabilities of PILOT to obtain a long
time-series of sharp images. The proposed 0.5~\degsq\ field-of-view of the PILOT PVISC
camera will allow us to capture millions of stars in \ocen, of which several tens of
thousands will oscillate with measurable amplitudes and periods. Our main targets will be the bright red giants, for which the predicted periods range from several hours to days.
To achieve our goals, we will need about one month of reasonably continuous observations
(target-of-opportunity breaks are acceptable). Observations will be principally in the
$r$-band, with exposures probing the range 13--16~mag. The high photometric precision
will allow us to probe less luminous stars, further down on the multiple red giant
branches.

Accurate stellar parameters from asteroseismology should enable us to
resolve the degeneracy between age, mass and helium abundance, and to test
directly whether more metal-rich stars are younger, as predicted by the
scenarios of prolonged star formation.  Comparing individual frequencies
of the various groups of stars in \ocen\ with stellar pulsation models
will also enable us to investigate the metallicity dependence of the oscillations.

Stars in \ocen\ can also be used to test predictions of stellar astrophysics, a
significant result alone added to the studies of the cluster as a system. As an added
bonus, \ocen\ contains thousands of classical pulsating stars, including large amplitude
RR Lyrae stars, Type II Cepheids, blue stragglers (SX Phoenicis stars), and eclipsing
binaries. PILOT will provide a unique data set on these objects, allowing us to put
further constraints on the properties of the cluster from analysing large numbers of
individual objects.

Finally, by co-adding the time-series photometry we would obtain the deepest and most
complete photometric catalogue of \ocen, extending down to the white dwarf cooling
sequence and the hydrogen burning limit (at $\mab\approx27$--28 in \vb). A $2\times2$
mosaic (four telescope pointings) coverage of the cluster will provide an unprecedented inventory of stars and produce an accurate measurement of the total cluster mass, a crucial parameter for interpreting the history and evolution of \ocen.

\subsubsection{Oscillating Stars in Open Clusters}

The southern open clusters NGC 3293 and NGC 4755 have ages around 10~Myr and present a
completely different population of stars compared to \ocen. Images of the clusters are
shown in Figure~\ref{astro_clusters}, indicating the field-of-view of PILOT, while
Figure~\ref{astro_cmd} shows the observed colour-magnitude diagrams. In this case, our
aim is not to study the cluster history, but rather to improve our understanding of
stellar physics. Stellar oscillations are standing sound waves, and a detailed comparison
of mode frequencies with stellar evolution models allows rigorous tests of theory. The
best studied oscillating star is the Sun, for which helioseismology has allowed
measurement of (among other things) the helium content, the depth the convection zone and
the internal differential rotation of the Sun.

With PILOT, we propose to apply asteroseismology to stars in these two open clusters.
This brings a huge advantage, since we can assume a common age, metallicity and reddening
for all the members of the cluster.  We can therefore study the slight differences from
star to star, and in this way make very stringent tests of stellar evolution theory.
Especially important are the hot main-sequence stars, some of which oscillate as
so-called \bcep\ variables.  Not to be confused with their better known and much cooler
cousins, the Cepheid variables, these \bcep\ stars have masses in the range 4--8~\msun\
and oscillate in multiple modes with periods in the range 4--8~hours. Asteroseismology of
\bcep\ stars is an exciting new field that is already providing strong input to
theoretical models, particularly regarding the important but poorly-understood processes
of rotation and internal mixing \citep{Pigulski_07}. Both clusters are already known to host
at least 10 \bcep\ variables, and in NGC 3293, one of them is in a detached eclipsing
binary, allowing a direct radius measurement \citep[$P \approx 8$ d;][]{Freyhammer_e_05}. A
detailed study of the \bcep\ stars in these clusters will provide an unrivalled set of
oscillation frequencies with which to confront theoretical models and improve our
understanding of the physics of massive stars.

This project is likely to require \si2~weeks of continuous observation per cluster
(again, target-of-opportunity breaks are acceptable). Observations will be primarily in
the \rb. The magnitude range of interest is 8--13, allowing this project to be
accomplished in bright sky (twilight/moonlight) conditions.

\section{Star and planet formation}
\subsection{Galactic Ecology --- Unveiling the \\Molecular Galaxy}
\subsubsection{Impact}

The site conditions on the summits of the Antarctic plateau makes possible sensitive
measurements in the mid-infrared of two of the lowest energy lines of molecular hydrogen.
This provides an opportunity for PILOT to directly map the principal component of the
molecular gas in the Galaxy, a project that no other existing or planned facility can
undertake.

Because the hydrogen molecule is so light, its rotational energy levels are well spaced,
with the lowest lying a few hundred degrees above the ground state. Moreover, these
transitions are emitted in a waveband poorly suited for observations from
temperate-latitude sites. Hence, the hydrogen molecule has been virtually impossible to
use as a probe of the molecular phase, unless in special conditions, such as those that
occur in shocks and ultraviolet-excitation, when the near-infrared vibrational lines are
excited in gas heated to \si1000~K \citep[e.g.,][]{Burton_92}.

The difficulties associated with observing hydrogen molecules directly has led to the use
of trace molecules, such as carbon monoxide, which emit in the millimetre wave bands, as
proxies for probing the molecular gas. These lines, however, are generally optically
thick, and the abundance and chemical state of the trace molecules is uncertain.

Direct measurement of the ground state molecular hydrogen lines would provide an
optically thin probe that samples the entire distribution. Our calculations show that
PILOT would be able to directly detect the emission from the 17~\mm\ S(1) and 12~\mm\
S(2) lines of \Hy\ in the typical warm environment of molecular clouds. Moreover, the
spatial resolution would be about 2$''$. This is more than an order of magnitude better
than achievable with mapping surveys using millimetre-wave telescopes and is nearly two
orders of magnitude better than the current best southern Galactic plane molecular
survey, conducted by the NANTEN telescope \citep{Mizuno_Fukui_04}. The ability to achieve
such a step forward in spatial resolution would revolutionise our view of the molecular
medium of the Galaxy.

The scientific context here is to understand the life history of the molecular phase of
the Galaxy---the processes by which molecular clouds are formed, go on to form stars and
are then dispersed. Yet both the formation of molecular clouds and the origin of their
physical state remain uncertain. Their masses are much larger than can be supported by
thermal pressure, giving rise to the concept of turbulent support. Our picture for the
molecular phase, of long-lived entities in quasi-equilibrium, has been giving way to one
of a much more dynamic environment, where the clouds and their internal structures may
actually be transient features \citep[e.g.,][]{Vazquez-Semadeni_e_06,Elmegreen_07}. In
the new picture, the clouds undergo dynamical evolution, being assembled rapidly by
supersonic compressions of the atomic medium, and become self-gravitating in the process.
The apparent equilibrium between the turbulent and gravitational pressures then simply
arises from the mode of rapid assembly. Star formation can then occur almost as soon as
the molecular clouds are formed.

This picture needs to be tested, through a comprehensive examination of the physical
environment across the range of spatial and density scales that exist in cloud complexes
and their surroundings, across the Galaxy. Central questions are: what is the turbulent
energy distribution? Does it relate to the contrasting pictures of local turbulence
injection from the natal stellar content (i.e., if virial equilibrium applies) or does it
arise from external sources (i.e., if compression applies)? These can be addressed by
unveiling the molecular galaxy, i.e., by mapping the distribution of it principal
tracer---the hydrogen molecule---on the arcsecond scale.

\subsubsection{The Mid-Infrared Lines of Molecular Hydrogen}

Most studies of the molecular hydrogen emission have so far focussed on the
near-infrared, where lines from shocks and photodissociation regions are prominent.
However, these do not mark the typical state of the molecular gas, only these regions
where vigorous activity is occurring, as the emission arises from energy levels several
thousand degrees above the ground state.

The lowest rotational lines of the hydrogen molecule, arising from the ground vibrational
level and emitted in the mid-infrared, allow the bulk of the gas to be accessed. The
ground state line itself, the 28~\mm\ 0-0 S(0) line, lies in a region of the spectrum
that can only be measured from space, but the 17~\mm\ 0-0 S(1) and 12~\mm\ 0-0 S(2) lines
may be accessed from Antarctica, where the background is an order of magnitude lower than
the best temperate sites, and the sky stability permits staring observations.

The bulk of the molecular gas is not at the \si10--20~K commonly associated with cold CO
emitting molecular clouds, but rather is at temperatures of order \si100~K. Much of this
exists in a photodissociation region (PDR) environment, warmed by the ambient
far-ultraviolet radiation fields to this temperature. These are the surface layers of
molecular clouds, but in fact comprise much of the molecular environment, outside the
cold, dense cores within the clouds where star formation is initiated. Furthermore, it
has also become clear through observations made with the ISO satellite that intensities
in the mid-infrared lines are significantly higher in many instances than would be
expected from simply PDR-excited molecular clouds \citep[e.g.,][]{Habart_e_05}.

Extensive regions of warm molecular gas ($T > 100$~K) exist throughout the cold
interstellar medium, for which ultraviolet photons cannot be the sole heating source.
This has led to the conjecture that turbulent heating may be significant. Turbulent
injection of energy and its dissipation is suspected of being the primary agent
responsible for the initiation and regulation of star formation through molecular clouds
\citep[e.g.,][]{Falgarone_e_05}. However, direct evidence for this is lacking because of
inability to probe the relevant molecular environment where this would occur. The warm,
molecular gas traced by the mid-infrared lines of molecular hydrogen are needed for this
task, as they are sensitive to the emission from weak shocks (e.g., $V_{shock}
\approx5$~\kms) where dissipation might occur.

There are even suggestions that in several external galaxies some baryonic dark matter
may be hidden inside such molecular gas, and has remained undetectable to date
\citep[e.g.,][]{Pfenniger_e_94}. The ability to trace the molecular medium in its primary
constituent, rather than in a trace one, would allow more reliable estimates to be made
of its contribution to the total baryonic mass content, and so assess the importance to
the dark matter budget.

The mid-infrared \Hy\ lines will also be sensitive to regions of activity in the
molecular gas, for instance where it is heated by shock waves from outflows from young
stellar objects, or by fluorescent emission from ultraviolet photons generated by young
stars nearby \citep[e.g.,][]{Burton_e_92}. These are generally much brighter than the
thermal emission from the quiescent gas, and so will be readily detected. Hence, the
proposed surveys will also be sensitive to the most active regions of molecular gas
across the Galaxy. While a dedicated survey for such active regions alone might be better
carried out through imaging the vibrational-rotational lines of \Hy\ in the 2~\mm\ band,
this information will be obtained at no extra cost through the survey we propose here.

\subsubsection{Other Facilities}

While the molecular hydrogen ground vibrational state S(1) and S(2) lines can be observed
from excellent observing sites such as Mauna Kea, there has been limited work undertaken
in this arena. The high background and unstable observing conditions preclude extensive
surveys, and there have been none proposed for such temperate sites.  Pointed,
spectroscopic observation of a few bright sources is all that has been done.  Airborne
observations, such as from SOFIA, might also access the S(1) and S(2) lines (but not the
28~\mm\ S(0) line); the only instruments proposed, however, are spectrometers, able to
obtain complete spectra across the relevant wavebands. It would not be possible to
conduct an imaging survey with the limited amount of observing time that would be
available from SOFIA.

Observations from space, with cryogenic instrumentation, are, of course, vastly more
sensitive than from even the best ground-based observatories operating in the thermal
infrared. However, the space observatories concentrate on deep photometric imaging in the
infrared bands, together with low spectral resolution spectroscopic capabilities for
pointed observations. For instance, JWST will be able to take spectra of the mid-infrared
\Hy\ lines in selected sources, but will not conduct imaging surveys of them. A proposal
for a suitable \Hy\ line imager, H2EX \citep{Boulanger_e_08}, was submitted to the ESA
Cosmic Visions program, but was not selected for further study.

There is thus no competition for PILOT for wide-field, spectroscopic, mid-infrared line
imaging surveys from any other facility. This provides a unique opportunity for the
telescope, should there also be scientifically worthy projects to conduct.  One such
project clearly stands out: imaging the molecular gas of our Galaxy in its primary
component. Such a facility would revolutionise our view of the molecular medium of the
Galaxy, and our understanding of the Galactic ecology.

\subsubsection{Observations}

We envisage a camera containing a wide-field Fabry-Perot, tuneable to the wavelengths of
interest, used in conjunction with a large format, mid-infrared array detector---this is
PMIRIS. With several months of dedicated telescope time, which could be in daylight as
this makes little difference in Antarctica, images of the molecular hydrogen emission
could be obtained across extensive regions of the Southern Galactic Plane.

\begin{figure}[t]
\begin{center}
\includegraphics[width=7.0cm]{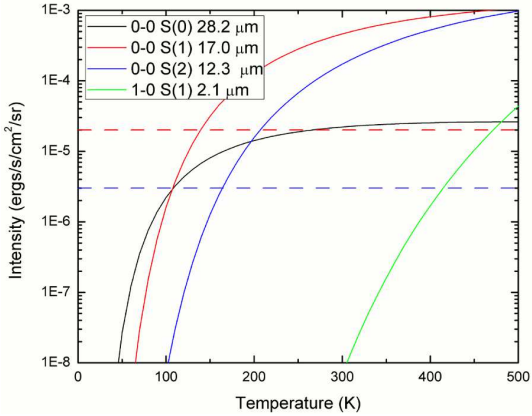}
\caption{Intensities of the molecular hydrogen emission lines from the mid-infrared 0-0
S(0), S(1) and S(2) lines (28.2, 17.0, 12.3~\mm), as well as the commonly observed
near-infrared 1-0 S(1) line (2.12~\mm), as a function of the gas temperature. These are
for a column density of gas equivalent to a unit optical depth (i.e., NH$_{2} \approx
10^{21}$~\cms), typical of the surface layer for a molecular cloud, and assume LTE. The
horizontal lines show the 1~$\sigma$ intensities that would be obtained in 10 minutes
integration time with PMIRIS for the S(1) and S(2) lines. As is readily apparent, the
near-infrared 1-0 S(1) line is unobservable in warm molecular gas. While the 0-0 S(0) line is
the most sensitive to warm molecular gas, down to \si100~K, it is unobservable from the
ground. The 0-0 S(1) and S(2) lines therefore provide the most ready access for directly
sampling the principal component of molecular clouds. Note that, while the sensitivity to
the S(2) line is greater, its higher energy level (1682~K c.f. 1015~K for the S(1) line)
means that the S(1) line is sensitive to cooler gas. It is emitted at a wavelength where
the Antarctic site is greatly superior to temperate-latitude observatories.}\label{H2_sens}
\end{center}
\end{figure}

PILOT would achieve a 1~$\sigma$ sensitivity in 10~minutes to the 17~\mm\ S(1) and
12~\mm\ S(2) lines of $2 \times 10^{-5}$ and $3 \times 10^{-6}$ \ergs,
respectively\footnote{Note that, while the 28~\mm\ 0-0 S(0) line is unobservable from any
Earth-based location, the 9.6~\mm\ 0-0 S(3) is also poorly situated for observation,
lying in a strong atmospheric ozone feature.}. These calculations assume background
limited operation, with background fluxes of 1000 and 30~Jy/arcsec$^{2}$ and atmospheric
transmissions of 0.7 and 0.98, at wavelengths of 17~\mm\ and 12~\mm, respectively (the
conditions are particularly favourable in Antarctica for observation of the 12 \mm\
line). They assume an instrumental efficiency of 0.3, a spectral resolution of 15\,000
($\Delta V = 20$~\kms), and 2$''$ pixel size (matched to the diffraction limit for the
17~\mm\ S(1) line). For well depths of $10^{6}$ electrons for the detectors, read out
would need to be every \si1~second and \si1~minute, for the two lines respectively; i.e.,
it would not be necessary to chop, but staring observations would suffice.

For a column density of molecular gas equivalent to an optical depth of unity (i.e.,
$A_{V} \approx 1 \equiv NH_{2} = 10^{21}$~cm$^{-2}$), these sensitivities equate to being
able to detect molecular gas warmer than \si140~K and \si165~K for the S(1) and S(2)
lines, respectively\footnote{For comparison, 1~minute integrations achieve sensitivities
and equivalent temperatures of $6 \times 10^{-5}$ \ergs\ \& 170~K and $10^{-5}$ \ergs\ \&
190~K, respectively for the S(1) and S(2) lines. 1~hour integrations will achieve $8
\times 10^{-6}$ \ergs\ \& 120~K and $10^{-6}$ \ergs\ \& 150 K. Note the cooler
temperatures reached by using the S(1) line, despite the poorer sensitivity at 17~\mm.
This is because the upper energy level is lower than that of the S(2) line.}---see
Figure~\ref{H2_sens}. Typical photodissociation regions have columns of $A_{v} \approx
3$--4 for their warm surface layers, so these sensitivities should be sufficient to pick
up gas to these limits in such environments.

These sensitivities should be compared to measurements already made of the lines to
demonstrate they are sufficient for the science needs. For instance, the Orion molecular
cloud shock exhibits line intensities of \si$5 \times 10^{-3}$ \ergs, when seen through
\si20$''$~beams \citep{Parmar_e_94,Burton_Haas_97}, and while the \Hy\ emission in this
source is an order of magnitude brighter than the typical Galactic source that is
studied, these fluxes are still another order of magnitude brighter than the sensitivity
limit.  Similarly, the photodissociation region in Orion has mid-infrared \Hy\ line
intensities of \si$3 \times 10^{-4}$ \ergs, an order of magnitude above the sensitivity
level \citep{Parmar_e_91}.  Over the Central Molecular Zone, the ISO satellite measured
intensities of \si$10^{-4}$ \ergs\ over large ($14''\times27''$) beams
\citep{Rodriguez-Fernandez_e_01} at a few selected positions. Even if the emission is
uniformly distributed over these beams, it would still be readily measurable with the
survey parameters above. Model predictions for a variety of molecular shock types show
that the mid-infrared lines will be readily detectable at the achievable sensitivity
levels, when the emission fills the pixel \citep{Burton_e_92}. They will also be
detectable in the photodissociation region environment when the ambient far-ultraviolet
radiation field, $G_{0}$, is $10^{4}$ or more times higher than the average interstellar
radiation field---as occurs when there is a nearby OB star.

The velocity spread of molecular hydrogen line emission across a typical Galactic source
would be no more than \si150~\kms. In the Central Molecular Zone, however, the emission
is spread over \si250~\kms. The wavelength change experienced by rays passing through the
plates at different angles when using a Fabry-Perot etalon, provides a means of also
measuring these motions. In order to image all the line emission across the field-of-view
of the detector, the plate separation needs to be scanned over the complete wavelength
range across the etalon. Alternatively, the etalon can be kept with a fixed spacing, and
the telescope moved in small steps (\si1$'$) across the array, so that each spatial
position is observed at many different phases across the complete wavelength range. Since
the observations would be background limited, a gain in sensitivity is achieved by using
higher spectral resolution (so long as the line is not resolved). Thus, the use of a
Fabry-Perot permits dynamical information on the source to be obtained in addition to the
imaging---through constructing a spatial-velocity cube of the line emission.

As an example of a possible survey that could be conducted, we consider a circular etalon
of diameter 15~cm, imaging over an array where the square aperture contained within the
etalon is 10$'$ on a side. Using a fixed etalon spacing, taking 1~minute of integration
per position, and stepping the telescope along a 1$'$ grid, each position in space would
be imaged 100~times, covering the phase velocity range of 175~\kms. For a Fabry-Perot
with spectral resolution 20~\kms, this yields \si10~minutes of integration per spectral
resolution element per position (the precise time depends on the velocity with respect to
that at the phase centre). To map a degree-sized field typical of a giant molecular cloud
would then require \si10~days of observing (assuming a 25\% duty cycle). To map the
complete $3\degs\times1\degs$ area of the Central Molecular Zone, which would also
require two separated spacings of the etalon plates in order to cover the \si250~\kms\
wide extent of the emission, would require \si60~days observing. In either case, the
observations would be sensitive to the presence of molecular gas warmer than \si140~K.
Such a project could be accomplished over a single summer observing period.

\begin{figure*}[t!]
\begin{center}
\includegraphics[width=14cm]{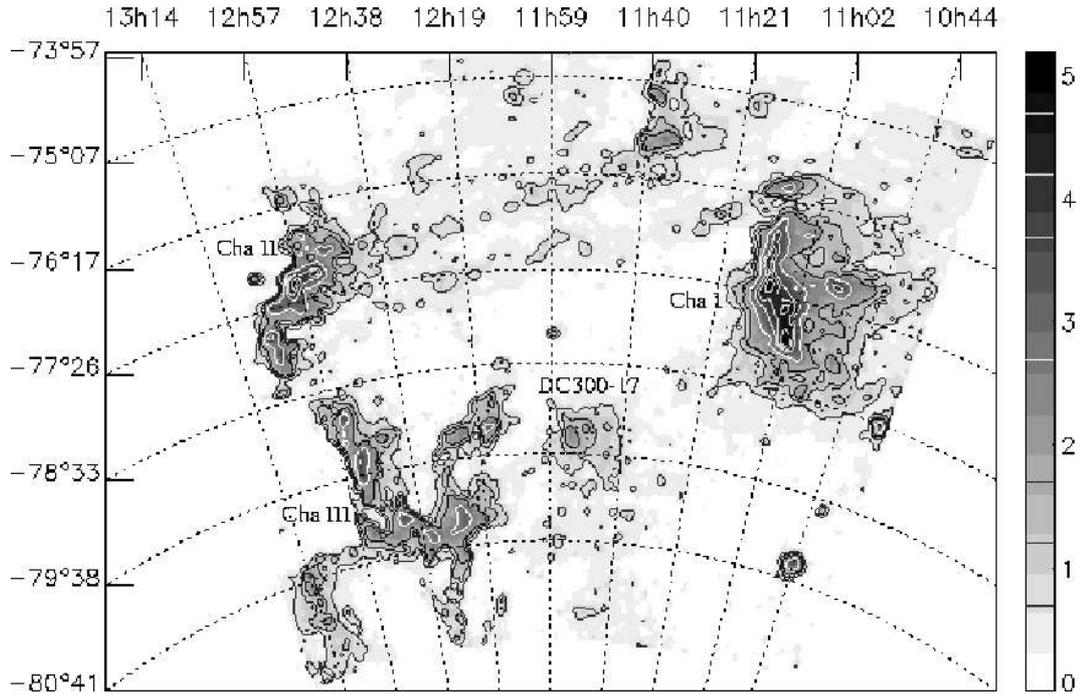}
\caption{Visual extinction map of the Chamaeleon dark clouds complex. From
\citet{Cambresy_99}.}\label{YSO_cham}
\end{center}
\end{figure*}

in addition to \Hy\, there are many other spectral lines in the mid-infrared bands that could also be imaged with a Fabry-Perot instrument. These are mostly fine structure atomic lines. Some are particularly strong coolants of the interstellar medium, for instance the [NeII] 12.8~\mm\ line, a dominant line emitted from H\textsc{II} regions. At 12.8~\mm\ the emission is nearly extinction free, so an imaging survey in the [Ne\textsc{II}] line would provide an unbiased view of H\textsc{II} regions across the Galaxy, providing dynamical as well as spatial information.  Such a survey could be conducted very much more quickly than the molecular hydrogen imaging survey.


\subsection{Disc Populations in the Cham-aeleon Dark Clouds Complex}

The studies of circumstellar discs specifically around young low-mass stellar objects
(YSOs) and young brown dwarfs (BDs) offer an opportunity to explore planet formation. The primordial circumstellar discs around low-mass stars and brown dwarfs can be easily
detected through observations at mid-infrared wavelengths longer than 5--10~\mm\ in which there is the best combination of contrast of the disc relative to the stellar
photosphere. In addition, measuring the lowest mass at which young objects harbour
circumstellar discs is crucial for determining whether planets can form around low-mass
BDs. In the last few years, thanks to the infrared satellites ISO and Spitzer, much
progress has been made in this subject. Using the IRAC and MIPS cameras on the Spitzer
telescope, large surveys of nearby star forming regions and young associations, such as
Taurus Perseus, Lupus, Serpens, and Chamaeleon \citep[e.g., see ][and references
therein]{Luhman_e_08a}, have identified  disc-bearing members at different masses and
stages of evolution.

The southern constellation of Chamaeleon (at a distance of \si160--180~pc) contains one
of the nearest groups of dark clouds to the Sun. This complex, shown in
Figure~\ref{YSO_cham}, is formed of at least six clouds as observed from the large scale
$^{12}$CO (J = 1-0) map of \citet{Mizuno_e_01} and the $A_{V}$ map of
\citet{Cambresy_99}. Chamaeleon was also the subject of investigation using the SPIREX
South Pole telescope \citep{Kenyon_Gomez_01}; 58\% of 124 sources detected in Cha~I at
\lb\ showed near-infrared excess emission characteristics of discs. Only small regions of the three main clouds (Cha~I, Cha~II, and Cha~III) have been extensively observed in the
mid-infrared with ISOCAM, the infrared camera of the ISO satellite
\citep{Persi_e_00,Persi_e_03}, and with Spitzer. The results of these surveys are well
summarized by \citet{Luhman_e_08b}. In an area of a few square degrees of Cha~I, 237
low-mass YSOs have been found, of which 33 have been identified as young BDs.
Approximately 50\% of these sources with masses from 0.01 to 0.3~\msun\ have discs. One
of the most exciting results is the discovery by \citet{Luhman_e_05} of a circumstellar disc in one of the less massive young BDs ($M\approx8$~$M_{J}$). About 50 members of Cha~II have been identified, including a few brown dwarfs \citep{Alcala_e_08}. In Cha~III, no sources with infrared excess have been found \citep{Persi_e_02}, suggesting that the cloud is probably at an early evolutionary stage of the star forming complex.

\begin{figure*}[t]
\begin{center}
\includegraphics[width=15cm]{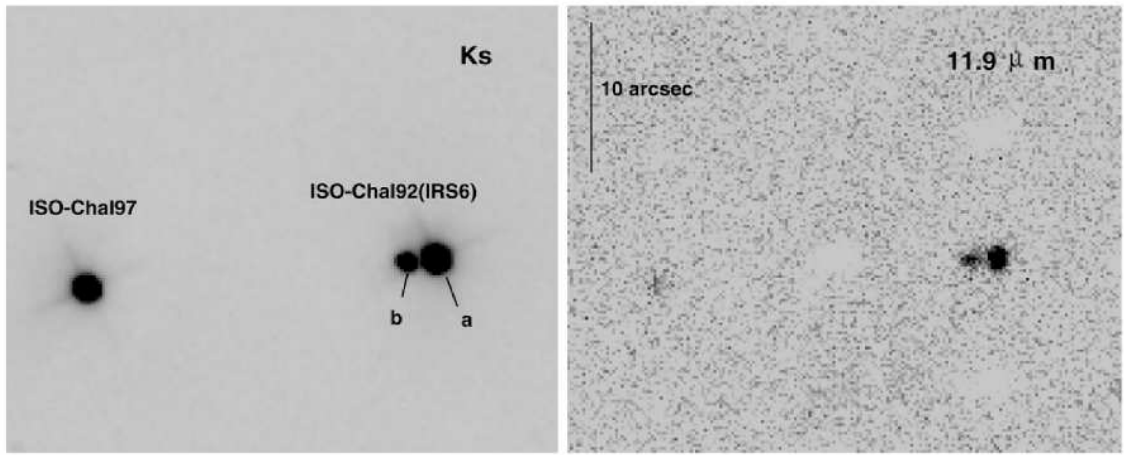}
\caption{$K_{s}$-band (left) and 11.9 \mm\ (right) images, obtained with the ESO 3.6~m New Technology Telescope, of the binary system ISO-ChaI92 composed of two Class I YSOs. Adapted from \citet{Persi_e_01}.}\label{YSO_2im}
\end{center}
\end{figure*}

The ISOCAM and Spitzer surveys of this region cover only a small fraction (5--8\%) of the total area (\si64~\degsq, see Figure~\ref{YSO_cham}). In addition, one of the limitations represented by these surveys in the mid-infrared is the poor spatial resolution (\si6$''$ FWHM at 24~\mm\ for Spitzer). This prevents the detection of multiple systems associated
with YSOs and young BDs, which are required to obtain a correct census of the young
stellar population in Chamaeleon that is fundamental to derive its initial mass function
(IMF). Figure~\ref{YSO_2im} shows, as an example, one of these multiple systems found in
Cha I by \citet{Persi_e_01}. The two YSOs have been classified as Class I sources
from their spectral energy distribution and are not separated in the Spitzer and ISOCAM
images.

The project proposed here for PILOT includes, as a first step, the survey of the whole
($8\degs\times8\degs$) Chamaeleon dark clouds complex in three different filters between
8 and 21~\mm. The required spatial resolution scales with the diffraction limit of the
PILOT telescope from 1$''$ (FWHM) at \si10~\mm\ to 1.8$''$ at 20~\mm; this is a factor 3
better than the MIPS resolution at 24~\mm. This will allow a complete census of YSOs and
young BDs from the analysis of the derived mid-infrared colour-colour plot. The
luminosities of these objects will be derived using \emph{J} and \emph{H} magnitudes
obtained from the 2MASS and VISTA surveys. It will thus be possible to obtain the most
complete IMF for this complex.

\begin{figure}[t]
\begin{center}
\includegraphics[width=7.5cm]{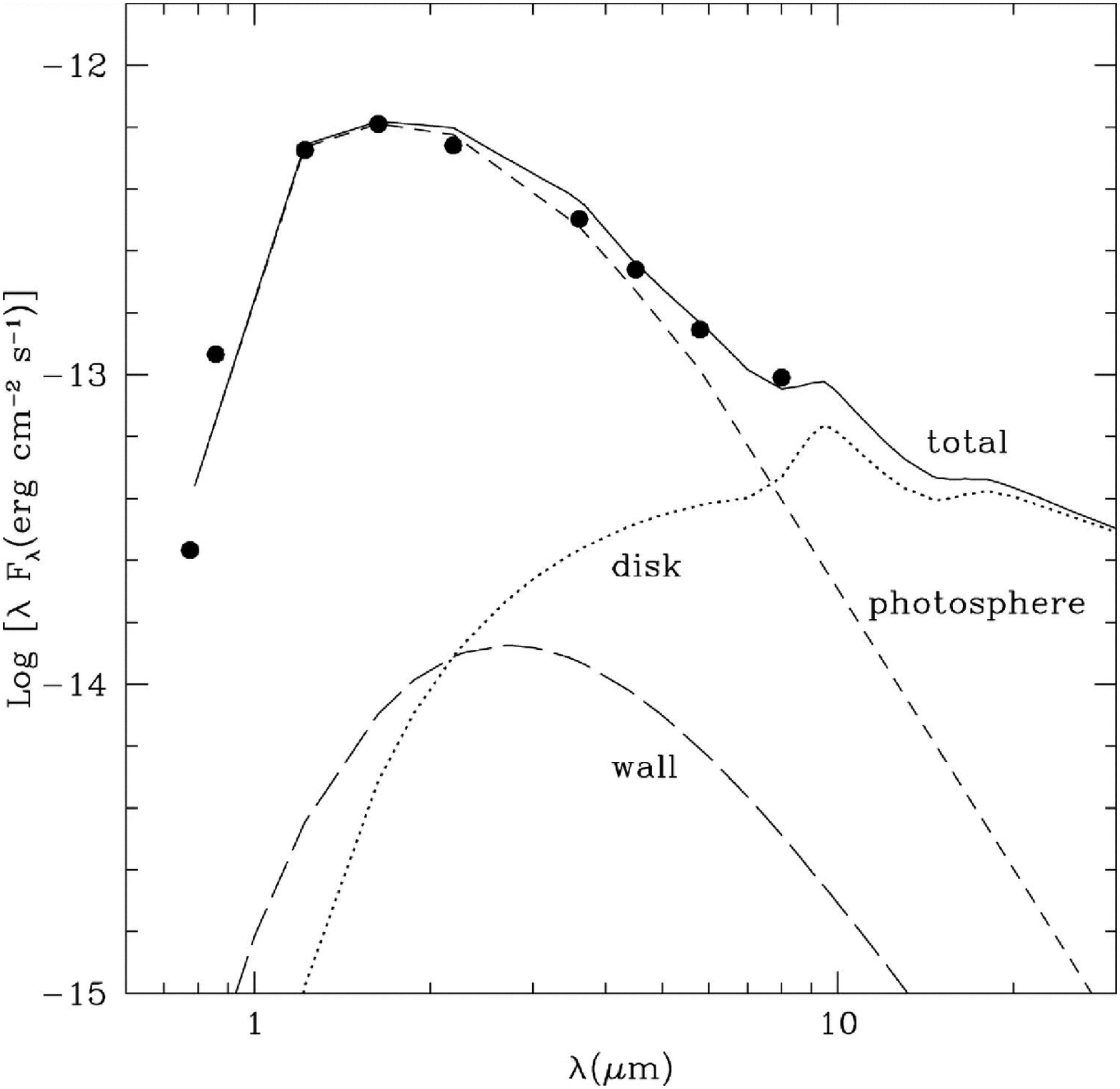}
\caption{Spectral energy distribution of the young brown dwarf Cha1109--7734 with mass
\si8 $M_{J}$. The observed excess flux at wavelengths greater than 5~\mm\ is modelled in
terms of emission from a circumstellar accretion disc. From \citet{Luhman_e_05}.}\label{YSO_SED}
\end{center}
\end{figure}

These observations can also be made using existing large telescopes (e.g., Gemini South,
VLT) equipped with mid-infrared imaging spectrometers (e.g., Michelle). However, such
facilities cannot be used to carry out large-scale mid-infrared surveys at reasonable
spatial resolution proposed here, because of their small field-of-view and the
instability of the 20~\mm\ window at mid-latitude locations. Current and near-future
infrared space missions such as AKARI and WISE will cover these wavelengths, but have a
poor spatial resolution and are thus not suitable to develop the proposed project.

As a second step of this project, photometry at longer wavelengths (30--40~\mm) of the
discovered YSOs and young BDs is required. This, together with the near-infrared and the
mid-infrared flux densities obtained in the survey, will allow spectral energy
distributions (i.e., as shown in Figure~\ref{YSO_SED}) to be built up over a wide range
of wavelengths. From comparison with existing models \citep[e.g.,][]{Robitaille_e_07}, it
will be possible to derive the physical characteristics of the circumstellar discs.

The baseline mid-infrared PILOT camera, PMIRIS, is suitable for this project. The large
survey will be made using the blue channel of this instrument with filters, chosen to
have a broad wavelength coverage, centred at 8.9, 12.7 and 21~\mm. With a $14'\times14'$
field-of-view, \si1400 frames per filter are necessary to survey the total area of the
Chamaeleon dark clouds, including an overlapping of 20\%. The survey depth should be on
the order of 5~mJy at 21~\mm, based on MIPS observations of Cha I and Cha II. This
represents an integration of \si1hr per frame at the PILOT $R = 20$ sensitivity, or
\si1400~hours total for the 21~\mm\ survey. Less time is required to obtain a similar depth at 8.9 and 12.7~\mm.. The total observing time for three filters is thus about 3000~hours. An extra time of \si500~hours will be necessary for photometry in one broad-band filter in the 30--40~\mm\ region using the red arm of PMIRIS for observations of young low flux YSOs and young BDs.

\subsection{Crystalline Silicates}

Dust grains in high temperature environments (e.g., protoplanetary discs and planetary
nebulae) undergo thermal processes which can be traced by the fraction of crystalline
silicates. The growth and crystallisation of sub-micron sized dust grains marks the onset
of planet formation in protoplanetary discs. Magnesium-rich compounds (forsterite,
enstatite) have spectral signatures in the mid-infrared, from 7 to 40~\mm.

\begin{figure}[t]
\begin{center}
\includegraphics[width=7.5cm]{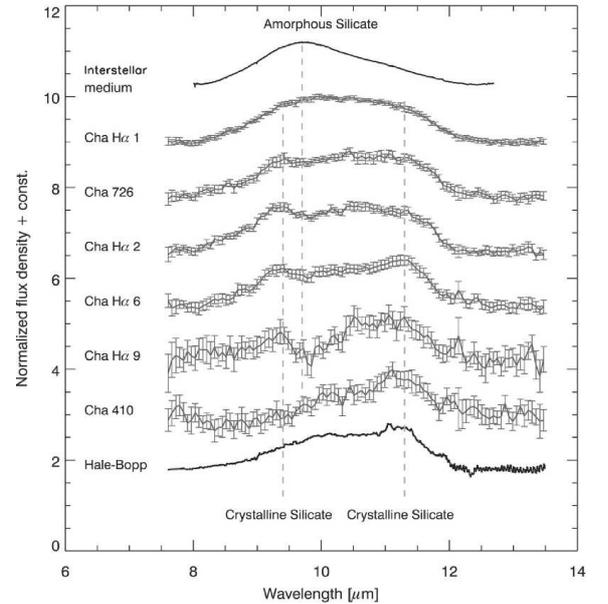}
\caption{Mid-infrared spectra of young BDs in Cha I obtained with the Spitzer Infrared
Spectrometer, compared with the spectra of the amorphous silicate-dominated interstellar
medium, and the crystalline-rich comet Hale-Bopp. From \citet{Apai_e_05}.}\label{sil_apai}
\end{center}
\end{figure}

\begin{figure}[t]
\begin{center}
\includegraphics[width=7.5cm]{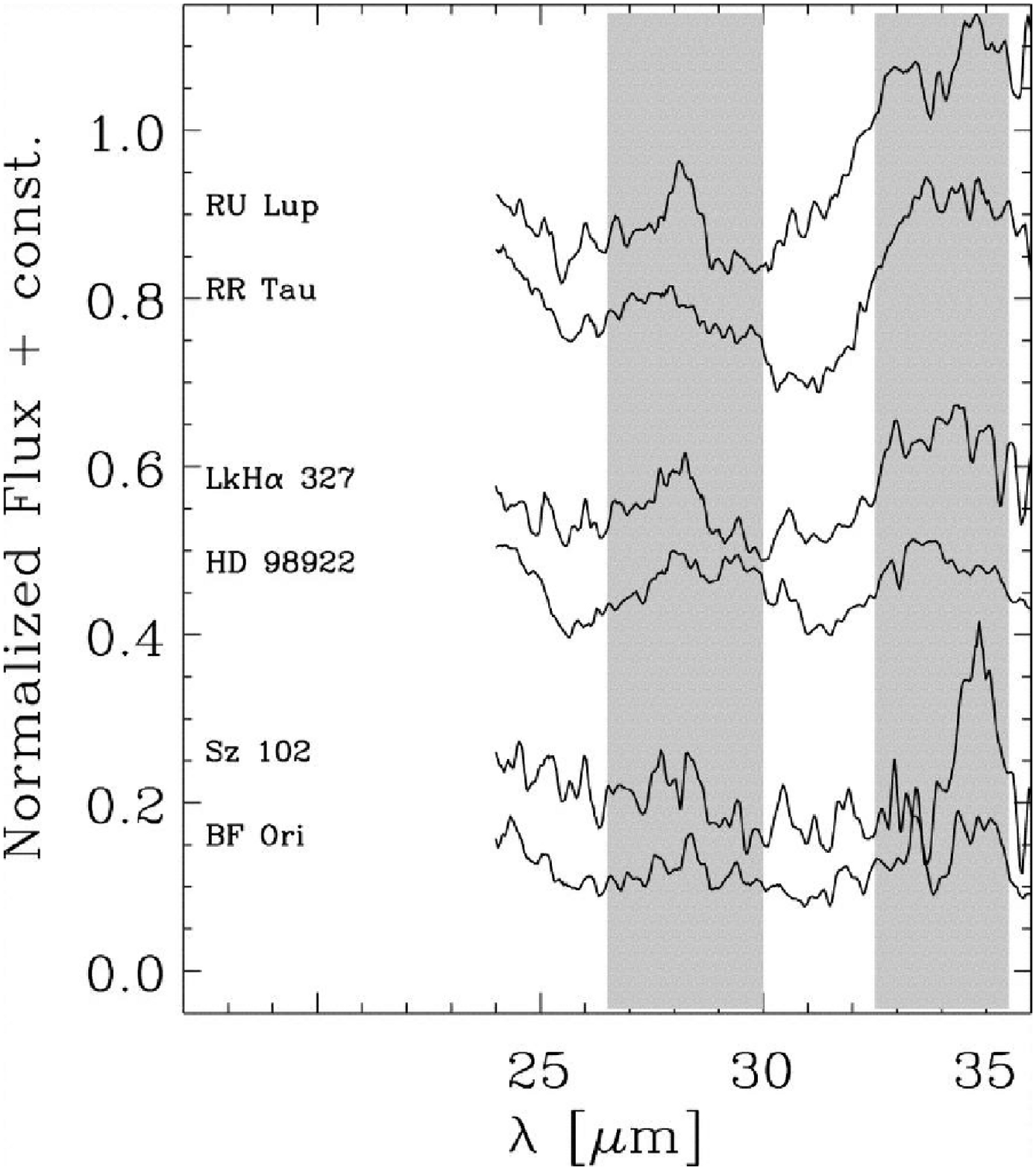}
\caption{A selection of spectra, after subtraction of disc emission, from sources showing
emission from crystalline silicates in the \si28 and \si34~\mm\ (shaded) regions. Each
pair of spectra contains one Herbig Ae/Be star (upper) and one T Tauri star (lower). A
high fraction of pre-main sequence stars (\si50\% for T Tauri stars) are found to exhibit
such emission features. From \citet{Kessler-Silacci_e_06}. }\label{sil_tauri}
\end{center}
\end{figure}

These features have been found in the 7--14~\mm\ wavelength range by \citet{Apai_e_05} in
a sample of young BDs in Cha I, as shown in Figure~\ref{sil_apai}. However, it is
difficult to disentangle the contributions from the organic molecules (PAHs), crystalline
silicates, and amorphous silicates in this wavelength range. Features at \si28~\mm\ and
\si34~\mm\ are only produced by crystalline silicates \citep[see ][]{Molster_Kemper_05}.
Figure~\ref{sil_tauri} shows this spectral range for several T Tauri and Herbig Ae/Be
stars. PILOT could study Mg compounds in nearby (\si200 pc) bright protoplanetary discs
around BDs and very low-mass YSOs.

While ISO/SWS and Spitzer/IRS have provided a small sample of low resolution spectra of
circumstellar discs in the wavelength range 25--40~\mm\ \citep[e.g.,][]{Morrow_e_08}, no
systematic survey has been undertaken or is currently foreseen. No existing or planned
ground-based telescope is sensitive in this wavelength range. The high wavelength cut-off
for the MIRI instrument\footnote{See \url{http://www.jwst.nasa.gov/miri.html}} on JWST is
27--29~\mm, the highest filter wavelength on WISE is 23~\mm, and the AKARI satellite only
observes shortward and longward of this range. The only future facility sensitive in this
region of the mid-infrared is the FORCAST instrument on SOFIA \citep{Adams_e_06}.
However, as one of a complement of 11 planned first generation instruments\footnote{See
\url{http://www.sofia.usra.edu/Science/instruments/sci_instruments.html}}, observing time
constraints will likely prevent it from performing large-area spectroscopic surveys. The
spectroscopic characterisation of a large sample of circumstellar discs, required to make
breakthrough advances in this field, can thus only be achieved with the PILOT telescope.

The project proposed here for PILOT is to perform a survey of the brightest nearby
protoplanetary discs in the 17--40~\mm\ wavelength range. Such a survey would allow the
thermal evolution of circumstellar discs to be traced, from the embedded to the pre-main
sequence and debris phases. This is a unique science case, because there are no other
facilities capable of performing such a systematic survey for Mg silicates in this
wavelength range.

This science case requires targeted low resolution ($R\approx500$) grism spectroscopic
observations in the range 17--40~\mm. This could be accomplished with the red arm of the
PMIRIS instrument. The estimated observing limit is for a bright protoplanetary disc at
up to \si200~pc for 1~hour of observing time ($\mab\approx 9$ at 30~\mm). No significant
temporal variability is expected, so the objects would only be observed once. The full
characterisation of all circumstellar discs bright enough to be observable from Dome~C is
a long-term project that will probably need more than one observing season. It is likely
that the brightest objects in the close Chamaeleon star forming region could be
completely observed during a single year, and the previously proposed Chamaeleon dark
clouds survey for circumstellar discs would provide the obvious source catalogue for such
a project.

\subsection{Embedded Young Stellar Objects}

\begin{figure*}[t]
\begin{center}
\includegraphics[width=15cm]{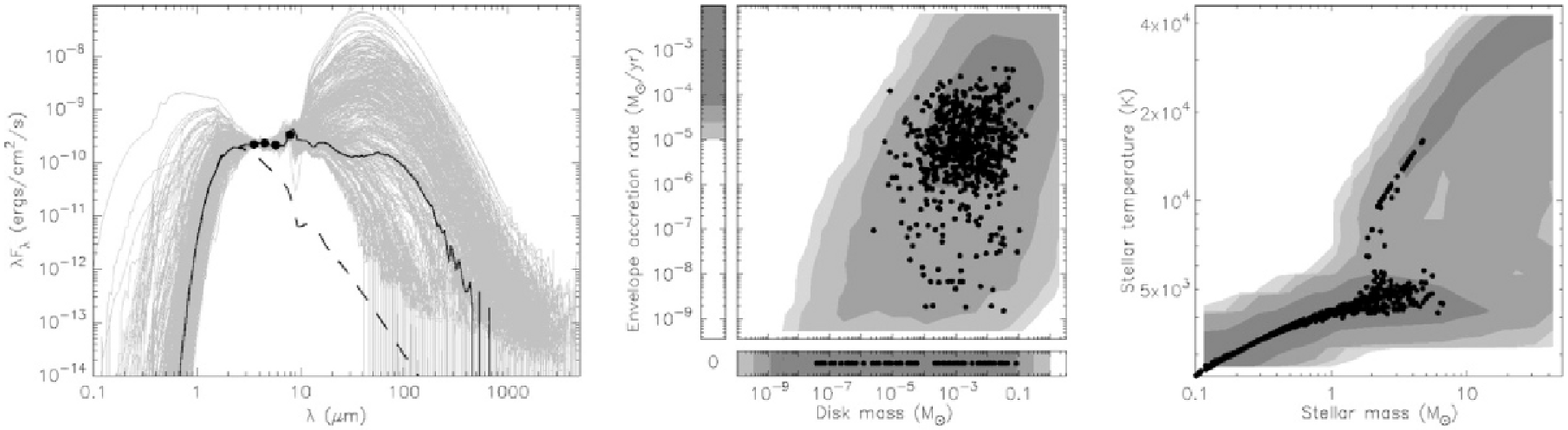}
\includegraphics[width=15cm]{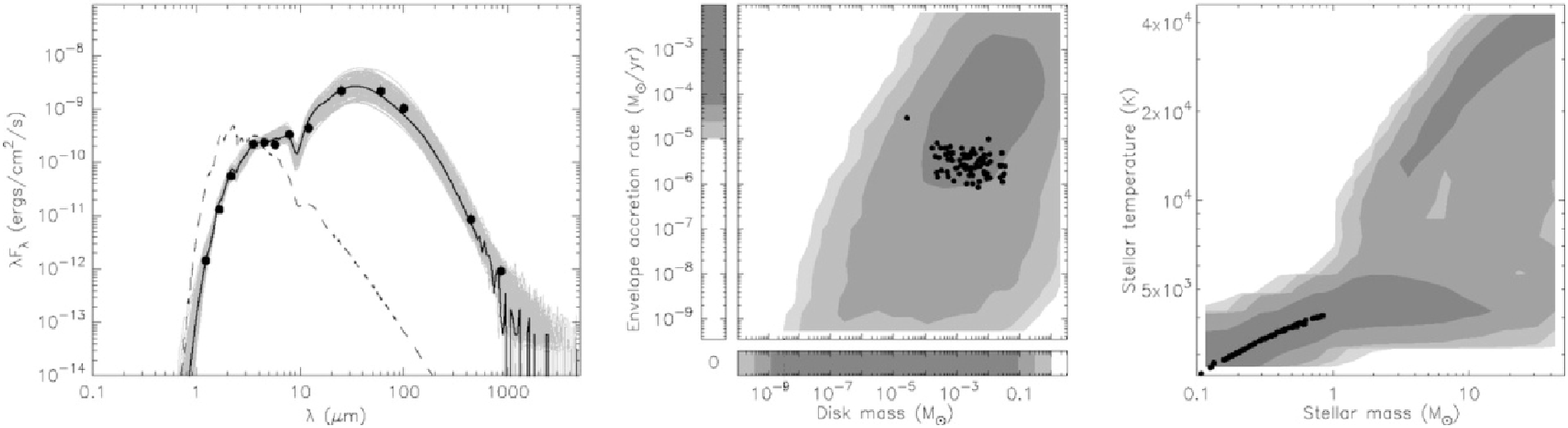}
\caption{The embedded object IRAS 04361+2547 spectral energy distribution (left column)
has been modelled using an increasing number of data points over a wider wavelength range
(top to bottom row). The mid- to far-infrared data effectively constrains the physical
properties of the object (centre and right columns). Adapted from
\citet{Robitaille_e_07}.}\label{EYSO_model}
\end{center}
\end{figure*}

Class 0/I Young Stellar Objects (YSOs, protostars surrounded by extended envelopes) are
cool objects (\si30--70~K). The emission maximum occurs at \si40--100~\mm. Therefore,
they can be best studied in the mid- to far-infrared, and space-based missions (IRAS,
ISO, Spitzer) have been essential to characterise and model those objects. Mid-infrared
fluxes can effectively constrain YSO models (see for example Figure~\ref{EYSO_model}) and
thus be used to estimate physical parameters such as stellar temperature, disc mass, and
accretion rate \citep{Robitaille_e_06,Robitaille_e_07}. Both high sensitivity and high
spatial resolution are required to break the many model degeneracies.

The space mission WISE is going to provide a global survey from 3.3 to 23~\mm\ with
unprecedented sensitivity but low spatial resolution. WISE will discover many embedded
YSOs and unresolved mid-infrared sources, but will not be able to fully characterise them
\citep{Mainzer_e_06}. PILOT could make follow-up observations in the spectral range
17--40~\mm\ with moderate spatial resolution, and provide key insight into the study of
YSOs.

The specific outcome of the proposed study is to obtain a photometric characterisation of
all the embedded YSOs observable from Dome~C in the 17--40~\mm\ range, and to obtain low
resolution spectroscopy of the brightest objects in the same interval. This project will
enable star formation to be studied in the less known very early stages, allowing
important new insights to be gained for embedded young stellar objects. Embedded YSOs
could be first studied in a systematic way in the wavelength range where many of them
have their maximum emission peaks, and ground-based temperate-location observatories have
zero atmospheric transmission.

Similar to the case for crystalline silicate observation, Spitzer/IRS has provided a
small sample of embedded sources with low resolution spectra in the wavelength range
25--40~\mm. No systematic photometric survey has been undertaken or is currently
foreseen. PILOT is the only facility (space or ground-based) apart from SOFIA/FORECAST
that will be capable of observing in this wavelength range.

The primary observing mode for this project is pointed observations of previously
discovered WISE embedded YSOs with the red arm of the PMIRIS instrument on PILOT.
Additionally, small surveys of molecular cloud cores may be undertaken. The estimated
observing limit is a solar mass embedded object at 5~kpc for 1~hour of observing time
($\mab = 12.2$ at 30~\mm). No temporal variability is expected, so the fields would only
be observed once.

Ideally, five filters would be used to accurately trace the spectral energy distribution
over the range 17--40~\mm. The full characterisation of all embedded protostars
observable from Dome~C will need more than one observing season. Nearby star forming
regions (e.g., Chamaeleon, Lupus) could be completely photometrically characterized
during a single year. Similar to the study of crystalline silicates, the proposed
7--25~\mm\ survey of the Chamaeleon dark clouds complex will provide many sources,
additional to WISE, for follow-up.

\section{Exoplanet Science}
\subsection{Planets ``Free Floating" in \\Nearby Star Clusters}

The detection of extra-solar planets orbiting other stars over the last decade has
revolutionised our understanding of how planets form and the amazing multiplicity of
extra-solar planetary systems. We have already found that solar systems that look just
like our own are not the norm, and may indeed not be all that common.

One of the primary difficulties in studying planets orbiting other stars is that it is
phenomenally difficult to access light that comes from the exoplanets themselves. Typical
contrast ratios between gas giant planets and their host star are on the order of
$10^{7}$ for very young exoplanets, and up to $10^{10}$ for Solar-age Jupiter-mass
planets in Jupiter-like orbits.

This makes it difficult to undertake detailed studies of the chemical and atmospheric
properties of exoplanets. Unfortunately, this is a critical gap in our understanding,
because at the low temperatures present in exoplanet atmospheres (from $\sim$200--1000~K)
the complexity of the molecular species that can exist makes constructing robust a priori
models for exoplanet spectra difficult. That is, while models can be constructed once we
know what an exoplanet's spectrum looks like, it is almost impossible to predict what
those spectra will look like in advance.

One way to make headway in understanding what the spectra of exoplanets will look like is
to seek locations where planets can be observed in the absence of a host star, by
searching for, and studying, objects of planetary mass ``free floating" in nearby stellar
clusters. We know from studies of several nearby star clusters that they host objects as
low in mass as a few Jupiter masses. Such objects are bright enough ($\mab\approx24$ at
\kb) for detailed study on 8~m and larger telescopes. PILOT therefore, has an exciting
opportunity to use its wide-field near-infrared survey capabilities to probe nearby star
clusters to identify their coolest ``T dwarf" members at temperatures of 500--1000~K.

A few investigators have searched for T dwarfs in younger (1--10~Myr) stellar
environments using traditional optical wide-field techniques \citep{Zapatero Osorio_e_02}
and narrow-field infrared techniques \citep{Najita_e_00,Lucas_Roche_00}. These have had
minimal success with only \citet{Zapatero Osorio_e_02} claiming the detection of a single
T dwarf. PILOT offers the opportunity to revolutionise this field---first because its
unprecedented image quality over wide fields is ideally suited to searching for faint
targets in the highly confused low galactic fields where star clusters lie. And secondly,
because it will be able to survey the large areas of sky that nearby star clusters
subtend (i.e., several \degsq) down to the magnitudes at which T dwarfs can be found
($mab\leq24$ at \kd-band). PILOT will be able to map an entire square degree in \emph{J},
\emph{H}, and \kd\ to the required depths in around 100~hours. Alternatively, with the
provision of suitable methane filters that can detect the unique spectral signature of
methane absorption in the \emph{H} or \kd\ bands for T dwarfs, a similar time would
enable deep direct methane imaging to the same depths, detecting objects as low in mass
as a few Jupiter masses, and enabling detailed follow-up studies of these ``planets
without stars".

The southern location of PILOT is an advantage for such work, as the southern sky hosts
most of the prime star cluster targets for free-floating planet work---including targets
such as R Corona Australis, Lupus, Chameleon, $\rho$ Oph, IC2391, and IC2602.

\subsection{Gravitational Microlensing}

The conditions at Dome~C provide opportunities for studies in planetary microlensing that
are not possible at any other site on the globe. By working in conjunction with the
existing OGLE-III and MOA-II telescopes in Chile and New Zealand that are dedicated to
gravitational microlensing, and also with the MicroFUN and PLANET networks of follow-up
telescopes, the PILOT telescope at Dome~C would provide the opportunity to make new
discoveries of ice-giant and terrestrial planets in the Galaxy. Approximately 150~hours
of telescope time on PILOT would be required per annum for 3 to 4~years.

Gravitational microlensing occurs if light rays from a distant star pass sufficiently
close to a nearer star so that their path is bent, or lensed, into pairs of images. These
images cannot be directly resolved but, as the lensing star moves across the line of
sight, the total magnified light from the background star can be measured and this
generates a symmetrical light curve profile that is now well recognised
\citep{Liebes_64,Paczynski_86}. If the lens star has a planetary companion, additional
lensing may occur, producing a perturbation in the light curve
\citep{Liebes_64,Mao_Paczynski_91}. This has now been demonstrated on a number of
occasions, some of which are referred to below. The first clear demonstration was carried
out by the OGLE and MOA groups, where an observed 7~day perturbation in the light curve
of a microlensing event was attributed to a 2.6~Jupiter mass planet in an orbit around a
0.63~solar mass main sequence star with an orbital radius of \si4.3~AU
\citep{Bond_e_04,Bennett_e_06}.

A particularly useful procedure for detecting planets in microlensing events was proposed
by \citet{Griest_Safizadeh_98}. They pointed out that if the lens star passes almost
directly in front of the background star, so that the magnification caused by lensing is
high, then planetary perturbations are very likely to occur at the peak of the event, and
these are likely to be of sufficient magnitude to be detectable. In these events of high
magnification, the pairs of images caused by lensing combine to form an almost complete
Einstein ring, whose radius is typically about 2--3~AU. \citet{Gaudi_e_98} pointed out
that multiple planetary perturbations could be detected in these events, leading to the
identification of relatively complex planetary systems. Furthermore,
\citet{Rattenbury_e_02} noted that the critical period for planet detection in these
events is the FWHM of the light curve, which typically persists for about 10~hours only.
Thus these events need only be intensively monitored over a brief period, during which
time they are relatively bright and easily observed. Finally, simulations reveal that
planetary perturbations in these events are surprisingly large. Ice-giant planets at
orbital radii lying within \si10\% of the Einstein radius yield perturbations of order
0.3~mag, easily detectable \citep{Yock_08}. Lighter planets, or planets further from the
Einstein ring, yield smaller perturbations, but nevertheless they may be detectable.
Heavier planets or planets closer to the ring, yield larger perturbations.

All of the above predictions have now been born out by observation. A planet of mass
3.4~\mjup\ was observed orbiting a star of mass \si0.46~\msun\ at \si3.5~AU, the most
massive planet observed orbiting an M dwarf star \citep{Udalski_e_05,Dong_e_08}. A planet
of mass \si13~$M_{\oplus}$ was detected orbiting a half solar mass star at \si2.7~AU
\citep{Gould_e_06}. This result, when combined with earlier results, led to the
conclusion that approximately one third of all stars have Neptune-like planets. A
planetary system was detected with planets of masses \si0.71 and \si0.27~\mjup\ at
orbital radii \si2.3 and \si4.6~AU orbiting a half solar mass star \citep{Gaudi_e_08}.
Most recently, a planet of mass \si3.3~$M_{\oplus}$ was detected orbiting a
\si0.06~\msun\ star that is likely to be a brown dwarf \citep{Bennett_e_08}. Other types
of observations have been made in microlensing events of high magnification. Stars have
been observed which do not have ice-giant or heavier planets at \si2--3~AU
\citep{Abe_e_04}. Also, the shapes of some stars have been measured
\citep{Rattenbury_e_05}. Additional examples of events like those described above are
presently under analysis.

However, despite the recent successes of microlensing, there have been a number of missed
opportunities where microlensers have been frustrated by poor weather and/or
interruptions by daylight at crucial times. PILOT would clearly be ideal for follow-up
observations of microlensing events; its location allows uninterrupted coverage of
selected events. Also the high cloud-free fraction allows a higher number of events to be
monitored and greatly reduces the number of missed opportunities. Furthermore, the seeing
is superb, enabling more precise measurements to be made, and the possibility exists for
all the critical observations in any one event to be made by a single telescope. This
would simplify the analysis of the data, and reduce systematic uncertainties. All the
above described successes by microlensing have occurred in observations made towards the
galactic bulge, and this presents a high air mass at Dome~C, with zenith angles ranging
from 44--80\degs\ \citep{Yock_06}.

A realistic strategy for Dome~C would be to concentrate on those events for which the
peak occurs at zenith angles $\leq$60\degs\ at Dome~C, and not to attempt to monitor
those at larger zenith angles. Nearly a half of all events should satisfy this criterion.
This would provide a sample of at least 30--40 fully-monitored events in a period of
3--4~years, as more than 20~events of high magnification are discovered annually by the
OGLE and MOA survey telescopes \citep{Yock_08}. The abundance of Neptune-like planets has
already been roughly determined from observations made at mid-southern latitudes
\citep{Gould_e_06}. With the enlarged sample that could be accurately monitored from
Dome~C, one could anticipate a firm determination of the abundance of ice-giant planets
being made in 3--4~years together with the first results on the abundance of terrestrial
planets. These measurements would complement corresponding measurements being made by the
radial velocity and transit techniques which enjoy greatest sensitivity to planets at
small orbital radii.

\subsection{Exoplanet Secondary Transits}

Recently a significant milestone in the field of exoplanets has been achieved: the direct
detection of infrared thermal emission of massive planets orbiting close to the star (the
so called ``hot Jupiters") via the observation of the secondary eclipse in the wavelength
range 3--20~\mm\ \citep[e.g.,][]{Charbonneau_e_08,Agol_e_05}. Up to now, all
positive detections have been made by the Spitzer satellite.

The observation of multiple exoplanet primary transits and secondary eclipses at infrared
wavelengths can allow the determination of high accuracy transit times, which could in
principle provide sensitivity to secondary lower-mass planets in resonant orbits
\citep{Agol_e_05}; high accuracy transit depths, which can be used to constrain the
planet molecular composition using atmospheric-absorption models \citep{Tinetti_e_07};
accurate eclipse depths, which can be used to determine the planet-to-star flux ratio,
and hence the planet thermal emission \citep{Charbonneau_e_08}; and variations of the
eclipse depth, which can be used to infer dynamical circulation regimes in the planet
atmospheres \citep{Rauscher_e_07}. To obtain such data requires a high temporal cadence
and a very stable infrared background; Dome~C is thus the only location on the ground at
which this science can be achieved.

The specific outcomes of the proposed study are to obtain high-precision photometric
light curves in the near and mid-infrared for the secondary transits of all the
transiting exoplanets detected in the southern hemisphere. The obtained spectral energy
distribution of these hot Jupiters will provide information about the upper atmospheric
layers of these planets, most notably the abundance of molecular species and the
turbulent mixing with lower layers. Such science thus allows the physical properties of
exoplanets to be probed.

Spitzer has provided high precision light curves for many transiting exoplanets, but the
detection of secondary transits from the ground has not yet been possible. The Spitzer
warm mission could provide 3.6 and 4.5~\mm\ broad-band photometric light curves for every
transiting exoplanet. However, no longer wavelength or spectroscopic facility is
available. JWST, with an exceptional sensitivity and low thermal background throughout
the thermal infrared, combined with a spectroscopic capability, will be an ideal tool for
this type of science. However, it is likely that alternatives to JWST will need to be
identified to undertake survey projects that require a large amount of observing time.

The primary observing mode for this project would be follow-up pointed observations of
previously discovered transiting exoplanets in \emph{L}, \emph{M}, and \emph{N} bands
with the PILOT PNIRC and PMIRIS cameras. The estimated observing limits for a hot Jupiter
around a F5V star are 360~pc (\lb), 190~pc (\mb) and 27~pc (\nb). For a K5V star, the
observing limits are 140~pc (\lb), 69~pc (\mb) and 10~pc (\nb). One or two secondary
transits would be observed per target. Additional intermediate-band filters would be
valuable to obtain a higher resolution spectral energy distribution. This
target-of-opportunity mode would follow the discovery announcements of major transit
survey projects and would be performed throughout the lifetime of the PILOT project.

\section{Solar system and Space Science}

\subsection{Planetary Imaging}

\begin{figure*}[t]
\begin{center}
\includegraphics[width=10cm]{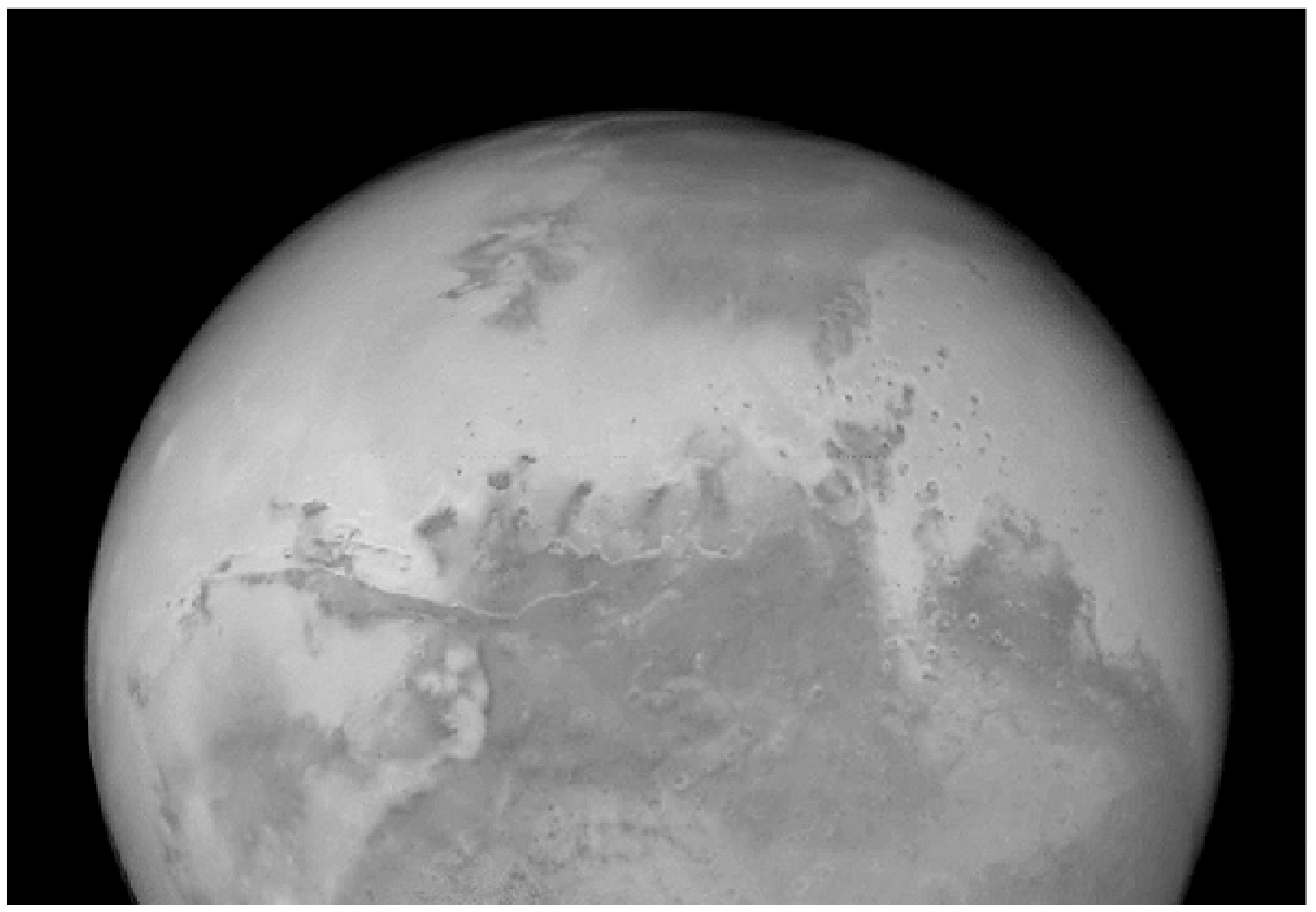}
\includegraphics[width=4.5cm]{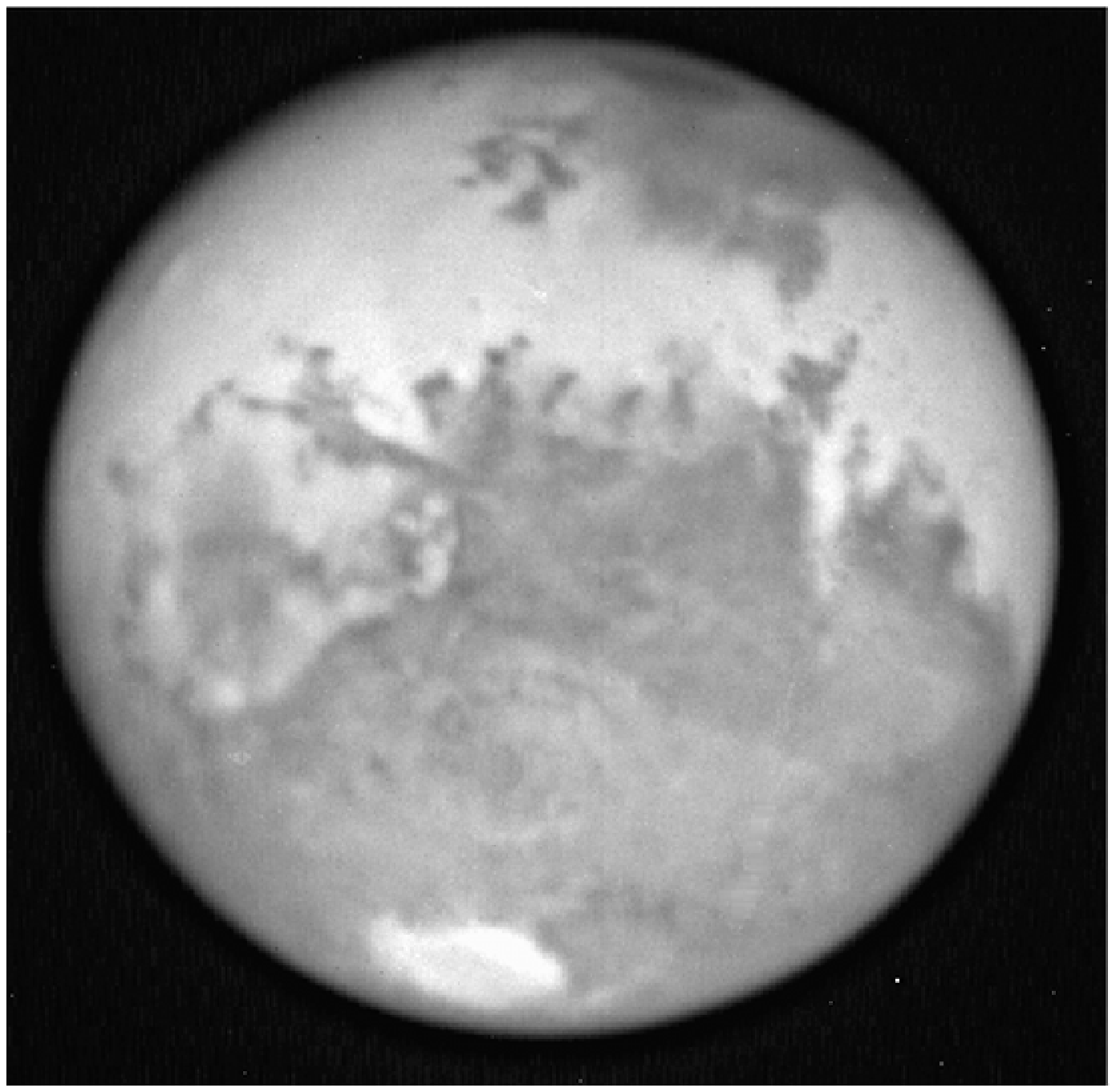}
\caption{Left: highest resolution image of Mars obtained with HST in Aug 2003. Right:
infrared image of Mars obtained at UKIRT in 0.35$''$ seeing using Lucky Imaging. PILOT
will improve on the UKIRT image at shorter wavelengths and should obtain images
comparable with that from the HST.}\label{planet_mars}
\end{center}
\end{figure*}

PILOT will be able to obtain exquisite images of planets with resolution comparable with
those achieved with the Hubble Space Telescope (see Figure~\ref{planet_mars}). While such imaging is possible for all
solar system objects the most valuable scientific application will be the observation of
the planets Mars and Venus. These objects cannot be imaged with current adaptive optics
systems due to the lack of suitable reference sources \citep{Bailey_04}. Venus cannot be
imaged with HST due to pointing constraints, while Mars was difficult to image in the
infrared with NICMOS due to saturation.

The Dome~C location means that the best observations will be restricted to times when the
planet is at its most southerly declination at opposition (Mars) or conjunction (Venus).
However, when these favourable opportunities occur, long durations of visibility are
possible. In the best cases, continuous 24-hour observations will be feasible. Venus will
have to be observed in daylight, but this should not be a problem owing to its brightness
and the reduced daylight sky brightness due to the low aerosol levels.

While these planets are also being studied by orbiting spacecraft, the ability of
ground-based observations to provide global continuous coverage provides data that is
complementary to the more detailed studies possible from spacecraft. The spatial
resolution achieved will be a good match to the resolution of atmospheric general
circulation models.

Specific studies, detailed below, that could be carried out on these two planets include
the atmospheric super-rotation of Venus, airglow in the upper atmosphere of Venus,
circulation and chemistry in the Venus lower atmosphere, and surface pressure on Mars.

The cloud layers of Venus show rotation periods of a few days, a much faster rotation
than that of the surface. The origin of this super-rotation is still not fully
understood, and is a current topic of much interest for atmospheric circulation modelers
\citep[e.g.,][]{Lebonnois_e_06}. Time-resolved imaging observations with high spatial
resolution can be used to measure the wind field by tracking the motion of cloud
features. Several levels in the Venus atmosphere can be studied, from 50~km using
nightside imaging at 1.7 and 2.3~\mm\ to around 70~km using wavelengths of 365 and
1000~nm on the dayside. While the ESA Venus Express spacecraft can also do this
\citep[e.g.,][]{Markiewicz_e_07}, its imaging observations are limited to the southern
hemisphere of Venus by orbital constraints. Current ground-based observations are
difficult to use for this purpose because of the limited duration of observations at any
one site.

Venus shows intense airglow emission in the singlet-delta band of molecular oxygen at
1.27~\mm\ that arises at an altitude of \si95~km. The intensity and spatial distribution
of this emission varies dramatically from night-to-night \citep{Crisp_e_96,Bailey_e_08}.
The airglow could easily be imaged with PILOT using an appropriate narrow-band filter,
and feature tracking could be used to study the wind field. The region appears to be a
chaotic one at the boundary between the retrograde super-rotation seen at the cloud tops,
and the sub-solar to anti-solar flow that dominates at high altitudes. Long-term
monitoring could be used to investigate whether the pattern swings between these two
states as has been suggested from sub-millimetre wind measurements \citep{Clancy_e_07}.
Once again PILOT can provide global coverage and continuous visibility not possible with
other spacecraft or ground-based systems.

The CO mixing ratio in the Venus lower atmosphere (\si30~km) can be measured from
nightside observations at around 2.3~\mm. It has been found to show latitudinal
variations with a peak at \si60\degs\ latitude and a minimum at the equator
\citep{Tsang_e_07,Marcq_e_06}. The distribution is believed to be the result of
photochemical formation of CO above the clouds, a circulation pattern that carries it
downwards at 60\degs\ latitude, and chemical removal of CO as it flows towards the
equator. While previous observations have used spectroscopy, modelling has shown that the
CO mixing ratio could be reliably determined from narrow-band filter imaging on and off
the 2.3~\mm\ CO band. PILOT could therefore be used to obtain long sequences of full-disc
images of the CO distribution that could be used to look for short- or long-term changes
in the CO distribution and hence the circulation patterns.

Observations of Mars in the CO$_{2}$ bands can be used to map the distribution of surface
atmospheric pressure \citep{Chamberlain_e_06}. While the main effect seen in such
observations is the surface topography, long series of observations of this type should
also reveal surface pressure effects due to weather systems. Surface pressure is not
measured routinely by any of the existing spacecraft. However, it is a key parameter
needed to help test and constrain general circulation models of the Martian atmosphere
\citep[e.g.,][]{Forget_e_99}.

The observations described above would be obtained with the PLEC and PNIRC instruments on
PILOT. Key requirements include the ability to obtain either short exposures and use
Lucky Imaging techniques, or a tip-tilt correction system capable of operating on the
image of a bright planet. For Venus, the telescope must be able to point close to the Sun
(at least down to 30\degs\ from the Sun). Additionally, the cameras must not saturate on
bright objects such as Mars and Venus (many astronomical cameras do). This either
requires the provision of sufficiently short exposure times, or availability of neutral
density filters.

\subsection{Solar Coronal Mass Ejections and Solar Flares}

Synchrotron radiations are emitted during solar flares and Sun coronal mass ejections and are usually observed in the microwave part of the spectrum. Understanding this phenomenon will shed light on the physical mechanisms that are responsible for solar flares and
coronal mass ejections, such as particle acceleration from the photosphere.

\begin{figure}[t]
\begin{center}
\includegraphics[width=7.5cm]{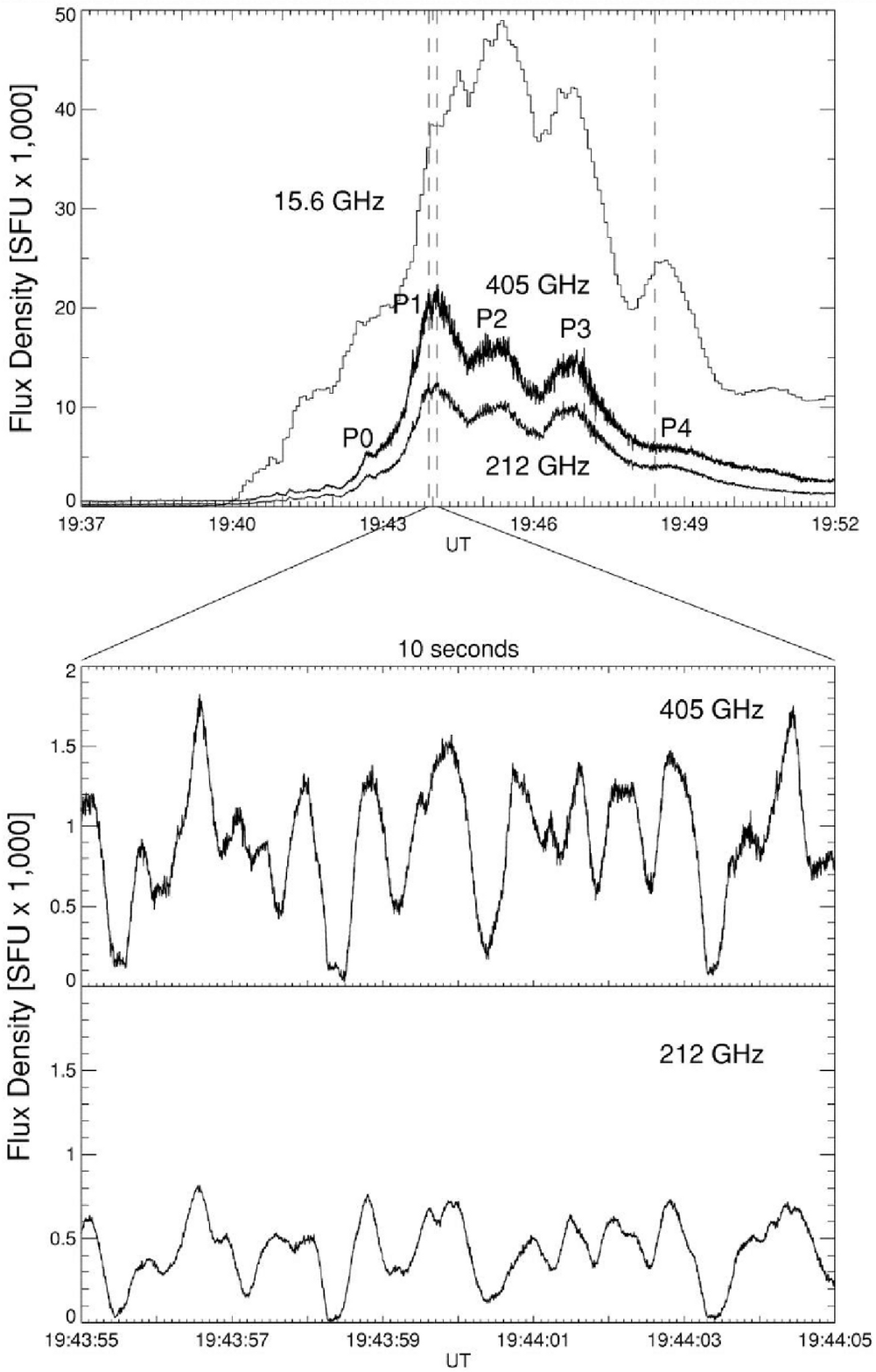}
\caption{Time profiles of the Solar burst in November 2003. The 405 GHz emission
is more intense than the 212 GHz emission, suggesting a possible maximum in the sub-millimetre/far-infrared domain.
From \citet{Kaufmann_e_04}. }\label{solar_CME_1}
\end{center}
\end{figure}

Recent results obtained at sub-millimetre wavelengths have revealed another potential
synchrotron radiation spectrum, as illustrated in Figure~\ref{solar_CME_1}
\citep{Kaufmann_e_04}. This new spectral component discovered with fluxes increasing for
shorter sub-millimetre wavelengths may indicate that the sub-millimetre emission is
created by particles accelerated to very high energies.

Current emission models assume three different mechanisms which may become comparable in
importance: (a) synchrotron radiation by beams of ultra-relativistic electrons, beams
different from the ones producing hard X-rays emissions (b) synchrotron radiation by
positrons produced by nuclear reactions arising from energetic beams interacting at dense regions close to the photosphere, and (c) free-free emission from the chromosphere
excited by high-energy electron beams producing hard X-rays emissions \citep{Kaufmann_e_06}.

Overall, these new results indicate that key questions regarding the physical mechanisms
at the origin of solar flares are expected to become better understood with measurements
in the far- to mid-infrared range. Sub-millimetre observations (at 200 and/or 350~\mm) of the Sun are proposed here using the SmilePILOT instrument. It could be used either in a
monitoring mode or as a follow-up imager if a flare is detected in the visible or
ultraviolet.

Current ground-based facilities capable of observing the Sun at sub-millimetre
wavelengths, such as the Solar Sub-millimetre Telescope \citep[SST;][]{Kaufmann_e_02}
located in Argentina, are limited by atmospheric conditions. The SST is a 1.5~m diameter
Cassegrain telescope that can observe at 212 GHz (1.5~mm) and 405 GHz (750~\mm). The FWHM
beams are 2$'$ and 4$'$ in these two respective channels. Observations with SST have
revealed sub-millimetre bursts in the Solar corona that were associated with large solar
flares, as shown in Figure~\ref{solar_CME_1}. Despite the large Solar flux density in the
sub-millimetre ($>$10~kJy), observations of the Sun from sites such as SST (at an
altitude of 2550~m) can only be done in the best atmospheric conditions when the
atmosphere may become opaque at these wavelengths. Observations at 200~\mm\ are basically
impossible from the SST site. It is important to obtain observations over a wide
wavelength range, from 500~\mm\ down to the mid-infrared, in order to understand whether
incoherent synchrotron radiations are the main mechanisms that generate sub-millimetre
flares. A second limitation of SST is that its small size does not allow high angular
resolution observations on the Sun ($<1'$).

Proposed space projects, such as the Small Explorer for Solar Eruptions (SMESE) project,
include instrumentation (the DESIR telescope) operating at both 150 and 35~\mm\
\citep{Millard_e_06}. However, this telescope is relatively small (\si170~mm), and the
SMESE bolometers will not be cryogenically cooled. Such a facility is thus limited in
terms of time, sensitivity, and resolution. Only major flares with respect to the thermal
emission of the Sun will be detected.

\begin{figure}[t]
\begin{center}
\includegraphics[width=7.5cm]{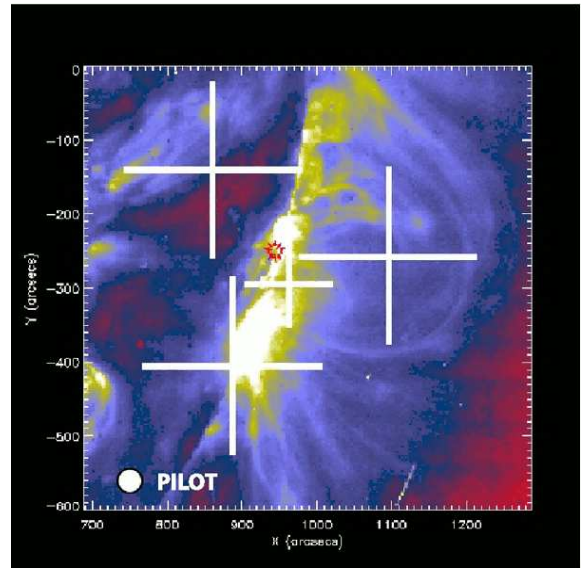}
\caption{Position and size of the sub-millimetre observing beam of the SST (crosses) on the SOHO
ultraviolet image of the Sun taken just before the November 2003 large flare, compared to
the size of the PILOT sub-millimetre beam (circle). Adapted from \citet{Kaufmann_e_04}.}\label{solar_CME_2}
\end{center}
\end{figure}

Observations of the Sun at Dome~C with PILOT can be accomplished at both 30~\mm\ and
200~\mm. Given the large flux of the Sun at these wavelengths, observations during the
daytime (summer months) with a precipitable water-vapour column density of 0.3~mm should
be possible. Cryogenically cooled detectors should allow detection of minor flares in
intensity. Additionally, PILOT would offer a better angular resolution to disentangle the
potential sources of flares on the Sun (see Figure~\ref{solar_CME_2}). A higher angular
resolution, and therefore a larger telescope than \si2.5 m in diameter, is not necessary
for observing the Sun.

Complementary observations in the mid-infrared with the PMIRIS instrument on PILOT would
also constrain the spectral energy distribution of the incoherent synchrotron radiation.
PILOT, as a solar corona mass ejection observer, will be highly complementary to the
Murchison Widefield Array (MWA) in Western Australia, which has a key science goal to
constrain the magnetic field strength during Sun coronal mass ejections
\citep{Salah_e_05}.

\subsection{Space Debris Tracking}

\begin{figure}[t]
\begin{center}
\includegraphics[width=7.5cm]{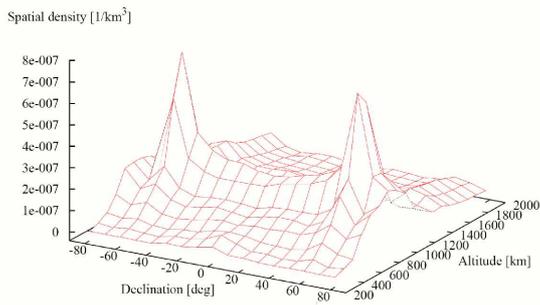}
\caption{The distribution of the spatial density of LEO debris as a function of altitude and
declination, based on the ESA MASTER'99 model for objects larger than 1~cm.}\label{LEO}
\end{center}
\end{figure}

There are estimated to be 120\,000 items\footnote{From the ESA Space Debris Mitigation
Handbook at \url{www.esa.int/gsp/completed/execsum00_N06.pdf}} of Low Earth Orbit (LEO)
debris larger than 1~cm and smaller than 10~cm that are not currently mapped (see
Figure~\ref{LEO}). This population poses a collision probability that represents a safety
risk for manned space-flight activities and a damage risk for spacecraft and satellites.
One of the ``advanced technology instrument" concepts for PILOT, the PILOT Satellite
Debris Camera (PSDC), contains a 1\degs\ field-of-view with a ring of
$1\mathrm{k}\times2$k CCD201 L3Vision detectors. This instrument would be capable of
identifying accurate orbits for most Low Earth Orbiting space debris larger than 1~cm
within a few years. This would occur at the cost of a small fraction of astronomical
observing time, if used in parallel with other instruments.

Sun-synchronous satellites have inclinations of 98.5\degs, and altitudes of 700--1000~km.
They can collide with debris in any orbit at the same altitude. Only debris with
inclinations 75--105\degs\ passes over Dome~C. For debris with a 100~minute orbit at
850~km altitude, the angular speed as it passes overhead is 1800~arcsecs per second; the
debris thus takes 66~ms to cross a CCD201 detector. Debris can only be seen when it is in
sunlight. At an 850~km altitude, this means a solar elevation greater than $-28\degs$.
Assuming the debris has surface brightness similar to the full moon, a 1~cm satellite at
850~km subtends 2.4~milli-arcsecs, giving a \vb\ magnitude $\mab=16.7$. This brightness
gives 530~photons on the detector for a crossing time of 66~ms. The debris track has a
FWHM width of \si0.2$''$ and is 120$''$ long. Assuming a grey sky with brightness $\mab =
19$ mag\,arcsec$^{-2}$, we get $\emph{SNR}\approx7$. Tracks with this \emph{SNR} would be
very obvious on Fourier transforming the images, and the double crossings provide
unambiguous confirmation.

Suppose we see a debris trail. Will we be able to catch it again, so as to fully
determine the orbit? If the debris is in sun-synchronous orbit, then it will pass roughly
overhead 24~hours later. It may be up to 50~minutes (half an orbit) early or late, during
which time the rotation of the earth will take Dome~C \si350~km out of the way, which is
a zenith distance of 23.5\degs\ for a height of 850~km. In 24~hours, the position of the
satellite is uncertain laterally by \si3~km or 12$'$~rms, and in the direction of motion
by 3~km~rms, or \si0.5~seconds in time. So we can predict the crossing 24~hours later
well enough to be sure to catch it.

The width of the PILOT field-of-view is 1\degs, or 14.8~km at 850~km altitude. The
circumference of the 75\degs\ S line of latitude is 10\,000~km. So, given that there are
two chances to cross the field-of-view on each orbit, there is a 0.3\% chance to see an
object (with inclination $>75$\degs) on each orbit. There are \si20\,000 such objects of
size $> 1$~cm within the appropriate declination and altitude ranges. The expected
discovery rate is then 36 per hour. This number could all be followed up 24~hours later.
So if PILOT were used to search for debris for an average of 100~minutes per day for a
year, it would (assuming Poisson statistics) see \si66\% of all such objects. A
systematic phase-space search pattern could probably be devised to do a much more
efficient search. Similar detection rates should be obtained for lower inclination orbits
($> 65\degs$) observed at higher zenith distance, as although the sky background is
higher (leading to a requirement for dark sky conditions) and the image quality worse,
the debris takes a longer time to cross the detectors. Thus, in principle, PILOT can
obtain orbits for most LEO debris, including almost all objects with inclinations
$>65\degs$, in a few years.

\section{Conclusion}

PILOT will be a powerful facility that will allow the detailed investigation of stellar
populations in nearby galaxies and stellar clusters. With a suite of instruments
comprising a high spatial resolution optical Lucky Imaging camera, a wide field-of-view
optical camera with moderate spatial resolution, and a high sensitivity near-infrared
camera, investigations of nearby galaxies at various distances are possible. Three
projects are proposed that will examine the Magellanic Cloud galaxies, a sample of Milky
Way satellite galaxies, and a sample of Local Group disc galaxies. Additionally, with a
stable point spread function and wide-field high-cadence capability, long time-series
asteroseismic observations of stellar clusters will be possible for the first time.

The mid-infrared capabilities of PILOT are uniquely suited to the study of star and
planet formation within the Galaxy. Firstly, a mid-infrared molecular hydrogen survey of
the Central Molecular Zone is proposed; this project, which aims to image a wide region
of sky with a high spatial resolution, is not possible with any other telescope.
Additionally, studies of circumstellar discs forming around early stage, and embedded,
young stellar objects and brown dwarfs, will be possible over wide areas with a high
spatial resolution and a broad wavelength coverage; these studies are only possible with
the PILOT telescope.

The detection and characterisation of exoplanets can also be addressed with PILOT. With a
wide field-of-view and high infrared sensitivity it should be possible to detect
free-floating planets in nearby stellar clusters to a low mass limit. The high-cadence
capability will allow the characterisation of gravitationally microlensed planets, with a
precision not possible with any other single telescope. Finally, a combination of high
infrared sensitivity, high cadence, and high atmospheric infrared background stability,
will allow the characterisation of transiting hot Jupiters, otherwise only possible with
space telescopes such as JWST and Spitzer.

Several projects have been proposed here for PILOT that involve observations within the
Solar System. Firstly, the Dome~C location, the Lucky Imaging optical camera, and the
wide-field near-infrared camera, will enable PILOT to obtain images of planets with a
resolution comparable to the Hubble Space Telescope, allowing details of the atmospheric
and surface conditions of Mars and Venus to be investigated. Additionally, the potential
sub-millimetre capabilities of PILOT will allow it to investigate the physics of solar
coronal mass ejections with a higher spatial resolution than other facilities. Finally,
if PILOT were equipped with a very wide-field fast imaging camera, its location would
allow it to determine the orbits for a large number of small-scale low earth orbit space
debris.

\section*{Acknowledgments}
The PILOT Science Case, presented here, was produced as part of the PILOT conceptual
design study, funded through the Australian Department of Education, Science, and
Training through the National Collaborative Research Infrastructure Strategy (NCRIS)
scheme, and the University of New South Wales through the UNSW PILOT Science Office. The
European contribution has been supported by the ARENA network of the European Commission
FP6 under contract RICA26150.


\begin{thebibliography}{}
\bibitem[Abe et al.(2004)]{Abe_e_04} Abe, F., et al. 2004, Science, 305, 1264
\bibitem[Adams et al.(2006)]{Adams_e_06} Adams, J.D., et al. 2006, Proc.~SPIE, 6269,
    626911
\bibitem[Agabi et al.(2006)]{Agabi_e_06} Agabi, A., et al. 2006, PASP, 118, 344
\bibitem[Agol et al.(2005)]{Agol_e_05} Agol, E., Steffen, J., Sari, R., \& Clarkson, W.
    2005, MNRAS, 359, 567
\bibitem[Alcal\`{a} et al.(2008)]{Alcala_e_08} Alcal\`{a}, J.M., Spezzi, L.,
    Chapman, N., \& Evans, N.J., et al. 2008, ApJ, 676, 427
\bibitem[Apai et al.(2005)]{Apai_e_05} Apai, D., et al. 2005, Science, 310, 834
\bibitem[Bailey et al.(2008)]{Bailey_e_08} Bailey, J., Meadows, V.S., Chamberlain, S., \&
    Crisp, D. 2008,	Icarus, 197, 247
\bibitem[Bailey(2004)]{Bailey_04} Bailey, J. 2004, PASP, 116, 745
\bibitem[Bekki et al.(2007)]{Bekki_e_07} Bekki, K., et al. 2007, MNRAS, 377, 215
\bibitem[Bennett et al.(2006)]{Bennett_e_06} Bennett, D.P., et al. 2006, ApJ, 647, L171
\bibitem[Bennett et al.(2008)]{Bennett_e_08} Bennett, D.P., et al. 2008, ApJ, preprint
    (arXiv:0806.0025)
\bibitem[Benson et al.(2007)]{Benson_e_07} Benson, A.J., Dzanovic, D., Frenk, C.S., \&
    Sharples, R.B. 2007, MNRAS, 379, 841
\bibitem[Bond et al.(2004)]{Bond_e_04} Bond, I.A., et al. 2004, ApJ, 606, L155
\bibitem[Boulanger et al.(2008)]{Boulanger_e_08} Boulanger, F., et al. 2008,
    Exp.~Astron., eprint(arXiv:0805.3109)
\bibitem[Bower et al.(2006)]{Bower_e_06} Bower, R.G., et al. 2006, MNRAS, 370, 645
\bibitem[Bullock \& Johnston(2005)]{Bullock_Johnston_05} Bullock, J.S., \& Johnston, K.V.
    2005, ApJ, 635, 931
\bibitem[Burton \& Haas(1997)]{Burton_Haas_97} Burton, M.G., \& Haas, M.R. 1997, A\&A,
    327, 309
\bibitem[Burton(1992)]{Burton_92} Burton, M.G. 1992, Aust.~J.~Phys., 45, 463
\bibitem[Burton et al.(1992)]{Burton_e_92} Burton, M.G., Hollenbach, D.J., \& Tielens,
    A.G.G.M. 1992, ApJ, 399, 563
\bibitem[Burton et al.(1994)]{Burton_e_94} Burton, M.G., et al. 1994, Proc.~ASA, 11,
    127
\bibitem[Burton et al.(2001)]{Burton_e_01} Burton, M.G., Storey, J.W.V. \& Ashley,
    M.C.B. 2001, PASA, 18, 158
\bibitem[Burton et al.(2005)]{Burton_e_05} Burton, M.G., et al. 2005, PASA, 22, 199
\bibitem[Cambr\`{e}sy(1999)]{Cambresy_99} Cambr\`{e}sy, L. 1999, A\&A, 345, 965
\bibitem[Chamberlain et al.(2006)]{Chamberlain_e_06} Chamberlain, S., Bailey, J.,
    \& Crisp, D. 2006, PASA, 23, 119
\bibitem[Charbonneau et al.(2008)]{Charbonneau_e_08} Charbonneau, D., Knutson, H.A.,
    Barman, T., Allen, L.E., Mayor, M., Megeath, S.T., Queloz, D., \& Udry, S., 2008,
    ApJ, submitted, eprint (arXiv:0802.0845)
\bibitem[Cioni et al.(2000)]{Cioni_e_00} Cioni, M.R., et al. 2000, A\&AS, 144, 235
\bibitem[Clancy et al.(2007)]{Clancy_e_07} Clancy, R.T., Sandor, B.J., \&
    Moriarty-Schieven, G.H. 2007, BAAS, 39, 539
\bibitem[Crisp et al.(1996)]{Crisp_e_96} Crisp, D., Meadows, V.S., B\'{e}zard, B.,
    de Bergh, C., Maillard, J.-P., \& Mills, F.P. 1996, J.~Geophys.~Res., 101, 4577
\bibitem[Demers et al.(2006)]{Demers_e_06} Demers, S., Battinelli, P., \& Artigau, E.
    2006, A\&A, 456, 905
\bibitem[Dong et al.(2008)]{Dong_e_08} Dong, S., et al. 2008, ApJ, submitted
    (Xiv:0804.1354)
\bibitem[Driver et al.(2007)]{Driver_e_07} Driver, S.P., Allen, P.D., Liske, J., \&
    Graham, A.W. 2007, ApJ, 657, L85
\bibitem[Eggen et al.(1962)]{Eggen_e_62} Eggen, O.J., Lynden-Bell, D., \& Sandage, A.R.
    1962, ApJ, 136, 748
\bibitem[Elmegreen(2007)]{Elmegreen_07} Elmegreen, B. 2007, ApJ, 668, 1064
\bibitem[Evans et al.(2005)]{Evans_e_05} Evans, C.J., et al. 2005, A\&A, 437, 467
\bibitem[Falgarone et al.(2005)]{Falgarone_e_05} Falgarone, E., et al. 2005, EAS~Pub.
    ~Ser., 14, 57
\bibitem[Ferguson et al.(2002)]{Ferguson_e_02} Ferguson, A.M.N., Irwin, M.J., Ibata,
    R.A., Lewis, G.F., \& Tanvir, N.R. 2002, AJ, 124, 1452
\bibitem[Forget et al.(1999)]{Forget_e_99} Forget, F., et al. 1999, J.~Geophys.~Res.,
    104, 24155
\bibitem[Freyhammer et al.(2005)]{Freyhammer_e_05} Freyhammer, L.M. et al. 2005, A\&A,
    429, 631
\bibitem[Fukugita et al.(1998)]{Fukugita_e_98} Fukugita, M., Hogan, C.J., \& Peebles,
    P.J.E. 1998, ApJ, 503, 518
\bibitem[Gaudi et al.(1998)]{Gaudi_e_98} Gaudi, B.S., et al. 1998, ApJ, 502, L33
\bibitem[Gaudi et al.(2008)]{Gaudi_e_08} Gaudi, B.S., et al. 2008, Science, 319, 927
\bibitem[Girardi et al.(2002)]{Girardi_e_02} Girardi, L., et al. 2002, A\&A, 391, 195
\bibitem[Gould et al.(2006)]{Gould_e_06} Gould, A., et al. 2006, ApJ, 644, L37
\bibitem[Griest \& Safizadeh(1998)]{Griest_Safizadeh_98} Griest, K., \& Safizadeh, N.
    1998, ApJ, 500, 37
\bibitem[Grundahl et al.(2006)]{Grundahl_e_06} Grundahl, F., et al. 2006,
    Mem.~Soc.~Astron.~Italiana, 77, 458
\bibitem[Gullieuszik et al.(2007)]{Gullieuszik_e_07} Gullieuszik, M., Held, E., Rizzi,
    L., Saviane, I., Momany, Y., Ortolani, S. 2007, A\&A, 467, 1025
\bibitem[Habart et al.(2005)]{Habart_e_05} Habart, E., et al. 2005, Space~Sci.~Rev., 119,
    71
\bibitem[Harris(1996)]{Harris_96} Harris, W.E. 1996, AJ, 112, 1487
\bibitem[Hidas et al.(2008)]{Hidas_e_08} Hidas, M.G., et al. 2008, Astron.~Nach., 329,
    269
\bibitem[Helmi(2008)]{Helmi_08} Helmi, A. 2008, A\&A~Rev., 15, 145
\bibitem[Holtzman et al.(2000)]{Holtzman_e_00} Holtzman, J.A., Smith, G.H., \& Grillmair,
    C. 2000, AJ, 120, 3060
\bibitem[Ibata et al.(1994)]{Ibata_e_94} Ibata, R.A., Gilmore, G., \& Irwin, M.J. 1994,
    Nature, 370, 194
\bibitem[Ibata et al.(2007)]{Ibata_e_07} Ibata, R., Martin, N. F., Irwin, M., Chapman, S., Ferguson, A.M.N., Lewis, G.F., \& McConnachie, A.W. 2007, ApJ, 671, 1591
\bibitem[Irwin et al.(2005)]{Irwin_e_05} Irwin, M.J., Ferguson, A.M.N., Ibata, R.A.,
    Lewis, G.F., \& Tanvir, N.R. 2005, ApJ, 628, L105
\bibitem[Kato et al.(2007)]{Kato_e_07} Kato, D., et al. 2007, PASJ, 59, 615
\bibitem[Kauffman et al.(1994)]{Kauffman_e_94} Kauffman, G., Guiderdoni, B., \& White,
    S.D.M. 1994, MNRAS, 267, 981
\bibitem[Kaufmann et al.(2002)]{Kaufmann_e_02} Kaufmann, P., et al. 2002, Rev. Mexicana
    Astron. Astrofis., 14, 149
\bibitem[Kaufmann et al.(2004)]{Kaufmann_e_04} Kaufmann, P., et al. 2004, ApJ, 2004, 603,
    L121
\bibitem[Kaufmann et al.(2006)]{Kaufmann_e_06} Kaufmann, P., et al. 2006, Solar Active
    Regions and 3D Magnetic Structure, 26th meeting of the IAU, JD03, 39
\bibitem[Kenyon \& G\'{o}mez(2006)]{Kenyon_Gomez_01} Kenyon, S.J., \& G\'{o}mez, M. 2001,
    ApJ, 121, 2673
\bibitem[Kenyon \& Storey (2006)]{Kenyon_Storey_06} Kenyon, S.L., \& Storey, J.W.V. 2006,
    PASP, 118, 489
\bibitem[Kenyon et al.(2006)]{Kenyon_e_06} Kenyon, S.L., Lawrence, J.S., Ashley, M.C.B.,
    Storey, J.W.V., Tokovinin, A., \& Fossat, E. 2006, PASP, 118, 924
\bibitem[Kessler-Silacci et al.(2006)]{Kessler-Silacci_e_06} Kessler-Silacci, J., et al.
    2006, ApJ, 639, 275
\bibitem[Kravtsov \& Gnedin(2005)]{Kravtsov_Gnedin_05} Kravtsov, A.V., \& Gnedin, O.Y.
    2005, ApJ, 623, 650
\bibitem[Kjeldsen et al. (2009)]{Kjeldsen_e_09} Kjeldsen, H., et al. 2009,
    IAU Symp. 253, 309
\bibitem[Lawrence(2004)]{Lawrence_04} Lawrence, J.S. 2004, PASP, 116, 482
\bibitem[Lawrence et al.(2004)]{Lawrence_e_04} Lawrence, J.S., Ashley, M.C.B.,
    Travouillon, T. \& Tokovinin, A. 2004, Nature, 431, 278
\bibitem[Lawrence et al.(2009a)]{Lawrence_e_09a} Lawrence, J.S., et al. 2009a, PASA,
    submitted (Paper I)
\bibitem[Lawrence et al.(2009b)]{Lawrence_e_09b} Lawrence, J.S., et al. 2009b, PASA,
    submitted (Paper II)
\bibitem[Layden \& Sarajedini(2000)]{Layden_Sarajedini_00} Layden, A.C., \& Sarajedini,
    A. 2000, AJ, 119, 1760
\bibitem[Lebonnois et al.(2006)]{Lebonnois_e_06} Lebonnois, S. et al. 2006,
    Super-rotation simulated with the new LMD Venus general Circulation Model, European
    Planetary Science Conference, Berlin, p167
\bibitem[Lee et al.(1999)]{Lee_e_99} Lee, Y.-W., et al. 1999, Nature, 402, 55
\bibitem[Liebes(1964)]{Liebes_64} Liebes, S. 1964, Phys.~Rev.~B, 133, 835
\bibitem[Lucas \& Roche(2000)]{Lucas_Roche_00} Lucas, P., \& Roche, P. 2000, MNRAS, 314,
    858
\bibitem[Luhman et al.(2005)]{Luhman_e_05} Luhman, K.L., et al. 2005, ApJ, 635, L93
\bibitem[Luhman et al.(2008a)]{Luhman_e_08a} Luhman, K.L., et al. 2008a, ApJ, 675, 1375
\bibitem[Luhman et al.(2008b)]{Luhman_e_08b} Luhman, K.L., et al. 2008b, ASP~Conf.~Ser.,
    Handbook of Star Forming Regions, in press
\bibitem[Maercker \& Burton(2005)]{Maercker_Burton_05} Maercker, M. \& Burton, M.G.,
    2005, A\&A, 438, 663
\bibitem[Mainzer et al.(2006)]{Mainzer_e_06} Mainzer, A.L., et al. 2006, Proc.~SPIE,
    6265, 626521
\bibitem[Majewski et al.(2005)]{Majewski_e_05} Majewski, S.R., et al. 2005, AJ, 130, 2677
\bibitem[Mao \& Paczynski(1991)]{Mao_Paczynski_91} Mao, S., \& Paczynski, B. 1991, ApJ,
    374, L37
\bibitem[Marcq et al.(2006)]{Marcq_e_06} Marcq, E., Encrenaz, T., B\'{e}zard, B., \&
    Birlan, M. 2006, Planet.~Space~Sci., 54, 1360
\bibitem[Markiewicz et al.(2007)]{Markiewicz_e_07} Markiewicz, W. et al. 2007, Nature,
    450, 633
\bibitem[Millard et al.(2006)]{Millard_e_06} Millard, A.A., et al. 2006, Proc.~SPIE,
    6266, 62660J
\bibitem[Mizuno \& Fukui(2004)]{Mizuno_Fukui_04} Mizuno, N., \& Fukui, Y., 2004,
    ASP~Conf.~Ser., 317, 59
\bibitem[Mizuno et al.(2001)]{Mizuno_e_01} Mizuno, A., et al. 2001, PASJ, 53, 1071
\bibitem[Molster \& Kemper(2005)]{Molster_Kemper_05} Molster, F., \& Kemper, C. 2005,
    Space~Sci.~Rev., 119, 3
\bibitem[Morrow et al.(2008)]{Morrow_e_08} Morrow, A.L., et al. 2008, ApJL, preprint
    (arXiv:0802.1732)
\bibitem[Mosser \& Aristidi (2007)]{Mosser_Aristidi_07} Mosser, B., \& Aristidi, E. 2007,
    PASP, 119, 127
\bibitem[Mosser et al.(2007)]{Mosser_e_07} Mosser, B., et al. 2007, EAS~Pub.~Ser., 25,
    239
\bibitem[Munoz et al.(2006)]{Munoz_e_06} Munoz, R.R., et al. 2006, ApJ, 649, 201
\bibitem[Najita et al.(2000)]{Najita_e_00} Najita, J., et al. 2000, ApJ, 541, 977
\bibitem[Nikolaiev et al.(2000)]{Nikolaiev_e_00} Nikolaiev, S., \& Weinberg, M.D.
    2000, ApJ, 542, 804
\bibitem[Olsen et al.(2003)]{Olsen_e_03} Olsen, K.A.G., Blum, R.D., \& Rigaut, F. 2003,
    AJ, 126, 452
\bibitem[Paczynski(1986)]{Paczynski_86} Paczynski, B. 1986, ApJ, 304, 1
\bibitem[Pancino et al.(2007)]{Pancino_e_07} Pancino, E., et al. 2007, ApJ, 661, L155
\bibitem[Parmar et al.(1991)]{Parmar_e_91} Parmar, P.S., Lacy, J.H., \& Achtermann, J.M.
    1991, ApJ, 372, L25
\bibitem[Parmar et al.(1994)]{Parmar_e_94} Parmar, P.S., Lacy, J.H., \& Achtermann, J.M.
    1994, ApJ, 430, 786
\bibitem[Persi et al.(2000)]{Persi_e_00} Persi, P., et al. 2000, A\&A, 357, 219
\bibitem[Persi et al.(2001)]{Persi_e_01} Persi, P., Marenzi, A.R., G\'{o}mez, M., \&
    Olofsson, G. 2001, A\&A, 376, 907
\bibitem[Persi et al.(2002)]{Persi_e_02} Persi, P., Marenzi, A.R., \& G\'{o}mez, M.
    2002, ESA~Pub.~Ser., 511, 221
\bibitem[Persi et al.(2003)]{Persi_e_03} Persi, P., Marenzi, A.R., G\'{o}mez, M., \&
    Olofsson, G. 2003, A\&A, 399, 995
\bibitem[Pfenniger et al.(1994)]{Pfenniger_e_94} Pfenniger, D., Combers, F., \& Martinet,
    L. 1994, A\&A, 285, 79
\bibitem[Pietrzynski et al.(2008)]{Pietrzynski_e_08} Pietrzynsky, G. et al. 2008, AJ,
    135, 1993
\bibitem[Pigulski(2007)]{Pigulski_07} Pigulski, A. 2007, Communications in
    Asteroseismology, 150, 159
\bibitem[Platais et al.(2003)]{Platais_e_03} Platais, I., et al. 2003, ApJ, 591, L127
\bibitem[Pohlen et al.(2008)]{Pohlen_e_08} Pohlen, M., Erwin, P., Trujillo, I., \&
    Beckman, J.E. 2008, ASP~Conf.~Ser., 390, 247 (arXiv:0706.3830)
\bibitem[Rattenbury et al.(2002)]{Rattenbury_e_02} Rattenbury, N.J., et al. 2002, MNRAS,
    333, 159
\bibitem[Rattenbury et al.(2005)]{Rattenbury_e_05} Rattenbury, N.J., et al. 2005, A\&A,
    439, 645
\bibitem[Rauer et al.(2008)]{Rauer_e_08} Rauer, H., Fruth, T., \& Erikson, A. 2008,
    PASP, 120, 852
\bibitem[Rauscher et al.(2007)]{Rauscher_e_07} Rauscher, E., Menou, K., Cho, J., Seager,
    S., \& Hansen, B.M.S. 2007, ApJ, 662, L115
\bibitem[Robertson et al.(2006)]{Robertson_e_06} Robertson, B., et al. 2006, ApJ, 645,
    986
\bibitem[Robitaille et al.(2006)]{Robitaille_e_06} Robitaille, T.P., Whitney, B.A.,
    Indebetouw, R., Wood, K., Denzmore, P. 2006, ApJS, 167, 256
\bibitem[Robitaille et al.(2007)]{Robitaille_e_07} Robitaille, T.P., Whitney, B.A.,
    Indebetouw, R., \& Wood, K. 2007, ApJS, 169, 328
\bibitem[Rodr\'{\i}guez-Fern\'{a}ndez et al.(2001)]{Rodriguez-Fernandez_e_01}
    Rodr\'{\i}guez-Fern\'{a}ndez, N.J., Martin-Pintado, J., Fuente, A., de Vicente, P.,
    Wilson, T.L., \& Huttemeister, S. 2001, A\&A, 365, 174
\bibitem[Salah et al.(2005)]{Salah_e_05} Salah, J.E., Lonsdale, C.J., Oberoi, D.,
    Cappallo, R.J., \& Kasper, J. 2005, Proc.~SPIE, 5901, 124
\bibitem[Saunders et al.(2008a)]{Saunders_e_08a} Saunders, W., Gillingham, P., McGrath,
    A., Haynes, R., Brzeski, J., Storey, J., \& Lawrence, J. 2008a, Proc.~SPIE, 7012, 70124F1
\bibitem[Saunders et al.(2008b)]{Saunders_e_08b} Saunders, W., Gillingham, P., McGrath,
    A., Haynes, R., Storey, J., Lawrence, J., Burton, M., \& Mora, A. 2008b, Proc.~SPIE, 7014, 70144N
\bibitem[Strassmeier et al.(2008)]{Strassmeier_e_08} Strassmeier, K.G., et al. 2008,
    A\&A, in press, preprint (arXiv:0807.2970)
\bibitem[Tinetti et al.(2007)]{Tinetti_e_07} Tinetti, G., et al. 2007, Nature, 448, 7150
\bibitem[Tomasi et al.(2006)]{Tomasi_e_06} Tomasi, C., et al. 2006, J.~Geophys.~Res.,
    111, D20305
\bibitem[Trinquet et al.(2008)]{Trinquet_e_08} Trinquet, H., Agabi, A., Vernin, J.,
    Azouit, M., Aristidi, E., \& Fossat, E. 2008, PASP, 120, 203
\bibitem[Tsang et al.(2007)]{Tsang_e_07} Tsang, C., et al. 2007, BAAS, 39,
\bibitem[Tully et al.(2005)]{Tully_87} Tully, R.B. 1987, ApJ, 321, 280
\bibitem[Udalski et al.(2005)]{Udalski_e_05} Udalski, A. 2005, ApJ 628, L105
\bibitem[Vazquez-Semadeni et al.(2006)]{Vazquez-Semadeni_e_06} Vazquez-Semadeni, E., et
    al. 2006, ApJ, 643, 245
\bibitem[Villanova et al.(2007)]{Villanova_e_07} Villanova, S., et al. 2007, ApJ, 663,
    296
\bibitem[Vlajic et al.(2008)]{Vlajic_e_08} Vlajic, et al. 2008, ApJ, submitted
\bibitem[Yock(2006)]{Yock_06} Yock, P. 2006, Acta~Astron.~Sin., 47, 410
\bibitem[Yock(2008)]{Yock_08} Yock, P. 2008, Proc. Manchester Microlensing Conference,
    preprint (arXiv:0805.1775)
\bibitem[Walden et al.(2005)]{Walden_e_05} Walden, V.P, Town, M.S., Halter, B., \&
    Storey, J.W.V. 2005, PASP, 117, 300
\bibitem[Whitney et al.(2008)]{Whitney_e_08} Whitney, B.A., et al. 2008, AJ, 136, 18
\bibitem[Worthey et al.(2005)]{Worthey_e_05} Worthey, G., España, A., MacArthur, L.A., \&
    Courteau, S. 2005, ApJ, 631, 82
\bibitem[Yong et al.(2006)]{Yong_e_06} Yong, D., Carney, B.W., Teixera de Almeida, M.L.,
    \& Pohl, B.L. 2006, AJ, 131, 2256
\bibitem[Zapatero Osorio et al.(2002)]{Zapatero Osorio_e_02} Zapatero Osorio, M.R., et
    al. 2002, ApJ, 578, 536
\end{thebibliography}
\end{document}